\documentclass[iop]{emulateapj}
\usepackage{epstopdf}
\shorttitle{Testing the dark energy with gravitational lensing statistics}
\shortauthors{Cao et al.}

\newcommand{\beq}{\begin{equation}}
\newcommand{\eeq}{\end{equation}}

\def\ba{\begin{eqnarray}}
\def\ea{\end{eqnarray}}

\begin{document}
\title{Testing the dark energy with gravitational lensing statistics}
\author{Shuo Cao\altaffilmark{1,2}, Giovanni Covone\altaffilmark{2,3}and Zong-Hong Zhu\altaffilmark{1}
}

\altaffiltext{1}{Department of Astronomy, Beijing Normal University,
100875, Beijing, China}
\altaffiltext{2}{Dipartimento di Scienze Fisiche, Universit\`a di Napoli ''Federico II'', Via Cinthia, I-80126 Napoli, Italy}
\altaffiltext{3}{INFN Sez. di Napoli, Compl. Univ. Monte S. Angelo, Via Cinthia, I-80126 Napoli, Italy}

\begin{abstract}

We study the redshift distribution of two samples of early-type gravitational lenses,
extracted from a  larger collection of 122 systems,
to constrain the cosmological constant in the $\Lambda$CDM model and
the parameters of a set
of alternative dark energy models (XCDM, Dvali-Gabadadze-Porrati and Ricci dark energy models), in a spatially flat universe.
The likelihood is maximized for $\Omega_\Lambda= 0.70\pm0.09$
when considering the sample excluding the Sloan Lens ACS systems (known to be biased toward
large image-separation lenses) and no-evolution,
and $\Omega_\Lambda= 0.81\pm 0.05$
when limiting to gravitational lenses with image separation $\Delta \theta > 2''$
and no-evolution. In both cases, results accounting for galaxy evolution are consistent within 1$\sigma$.
The present test supports the accelerated expansion, by excluding the null hypothesis
(i.e., $\Omega_\Lambda = 0 $) at more than $4 \sigma$,
regardless of the chosen sample and assumptions on the galaxy evolution.
A comparison between competitive world models is performed by means
of the Bayesian information criterion.
This shows that the simplest cosmological constant model that has only one
free parameter is still preferred by the available data on the redshift distribution of gravitational lenses.
We perform an analysis of the possible systematic effects, finding that the
systematic errors due to sample incompleteness, galaxy evolution and model uncertainties
approximately equal the statistical errors, with present-day data.
We find that the largest sources of systemic errors are the dynamical normalization and
the high-velocity cutoff factor, followed by the faint-end slope of the velocity dispersion function.

\end{abstract}

\section{Introduction}

In the last 15 years, several complementary observational probes on cosmological scales
have found strong evidence for an accelerating expansion of the universe:
distance measurements of distant Type Ia supernovae (SNe Ia; \citet{Riess98,Perlmutter99}),
the observations of the cosmic microwave background anisotropies (WMAP; \cite{Bennett03}), and
the baryon acoustic oscillations (BAOs) in the power spectrum of matter
extracted from galaxy catalogs \citep{Percival07}.
By assuming General Relativity, a negative pressure component has been
invoked as the most suitable mechanism for the observed acceleration, the simplest way of which is
the cosmological constant $\Lambda$, with a constant equation-of-state (EoS) parameter $w=-1$.
Dynamical dark energy models have also been proposed in the literature, such as the quintessence
\citep{Ratra88,Caldwell98}, phantom
\citep{Caldwell02}, quintom \citep{Feng05,Feng06,Guo05},
Dvali-Gabadadze-Porrati (DGP) \citep{Dvali00,Zhu05,Zhu08},
and Ricci dark energy (RDE) models \citep{Gao09,Li10}.

The existence of a large number of theoretical models has triggered a variety of observational tests,
based, for instance, on the angular size-redshift data of compact radio sources \citep{Alcaniz02,Zhu02},
the age-redshift relation \citep{Alcaniz03}, the lookback time to galaxy clusters \citep{Pires06},
X-ray luminosities of galaxy clusters, and the Hubble parameter data \citep{hz1,hz2,Cao11a,Cao11b}.
In this context, strong gravitational lensing plays an important role,
providing cosmological tests, such as gravitational lensing statistics
\citep{Kochanek96a,Zhu98,Cooray99,Chiba99,Chae02,Sereno05,Biesiada10,Cao11c,Cao12a,Cao12b},
Einstein rings in galaxy-quasar systems \citep{Yamamoto01},
clusters of galaxies acting as lenses on background high redshift galaxies \citep{Sereno02,Sereno04},
and time delay measurements \citep{Schechter04}.
Results from techniques based on gravitational lensing  are complementary to
other methods and can provide restrictive limits on the cosmological parameters.
In this paper we focus on one interesting lensing statistic suggested by \citet{Kochanek92}
and further discussed and developed in literature (e.g., \citet{Helbig96}, \citet{Ofek03}).

\citet{Fukugita90} (but see also \citet{Nemiroff89}) demonstrated that the gravitational lens expected redshift  increases
with a larger cosmological constant, $\Omega_{\Lambda}$. \citet{Kochanek92} derived the probability distribution of the redshift of the lens,
as a function of the cosmological parameters, at fixed image separation and source redshift. 
His analysis of a small sample of four lenses strongly favored a null cosmological constant with respect to models with large
$\Omega_{\Lambda}$ in a flat cosmological model. \citet{Helbig96} further discussed the method,
assuming  no evolution of the galaxy population
and no detection of  lensing galaxies beyond a given magnitude.
Successively, \citet{Kochanek96a} extended his analysis to include
gravitational lenses with not measured redshift,
in order to account for the selection effects.
He obtained the upper limit  $\Omega_{\Lambda}<0.9$
at the $95\%$ CL, assuming a flat cosmological model.
\citet{Ofek03} presented a  new analysis of expected lens redshift distribution,
taking into account number and mass evolution of the lens population, and applied this method to constrain
both the cosmological and mass-evolution parameter spaces.
They could obtain a strong upper limit on the cosmological constant ($\Omega_{\Lambda}<0.89$ at the $95\%$ CL),
in a flat universe with no lens evolution.

Compared with other related methods based on the full lensing probability distribution, all of the uncertainties in the absolute value of the optical depth can be eliminated in the differential probability of \citet{Kochanek92}, since it is determined by integrating the full lensing probability distribution over lens redshifts. Moreover, the sharp cutoff of this relative probability at high redshift makes the quantity $d\tau / dz_{\rm l} (\Delta\theta,z_{s})$ a more powerful cosmological probe \citep{Ofek03}.

Following works investigating the evolving lens population have
concluded that galaxy evolution is not strongly constrained by the redshift
distribution test and does not significantly affect lensing statistics \citep{Mit05,CN07,Oguri12}.
However, the evolution of mass and number density
can introduce large statistical errors and bias in the analysis of the lens redshift distribution.
Therefore, it is mandatory to consider the mass and density evolution into the statistical analysis of the redshift distribution of gravitational lenses. Other limitations include systematic effects due to a sample of gravitational lenses which completeness
might be not homogenous as a function of the lensed image separation and
the lens redshift \citep{CN07}.

\citet{Oguri12} have presented a comprehensive statistical analysis of the sample of 19 lensed quasars found
in the Sloan Digital Sky Survey (SDSS) Quasar Lens Survey (SQLS).
This sample is used to determine both the cosmological constant and evolution of the massive lensing galaxies.
When considering a no-evolution case, a null cosmological constant is rejected at 6$\sigma$ level,
providing an independent evidence for the accelerated expansion.

The purpose of this paper is to extend our previous statistical analysis
based on the angular separation distribution of the lensed images \citep{Cao11c}
by using the redshift distribution test to obtain novel constraints
on the parameters of spatially flat cosmological models.
With this aim, we use a large sample of 122 gravitational lenses
drawn from the Sloan Lens ACS (SLACS) Survey and other sky surveys.

The first aim is to obtain new constraints on the cosmological constant, by
assessing both statistical and systematic uncertainties, mainly due to galaxy evolution and sample selection.
Then, we also compare a number of alternative dark energy models with different
numbers of parameters, in our analysis we apply, following \citet{Davis07, Li10}, a model comparison statistic, i.e.,
the so-called Bayesian information criterion (BIC; \citep{Schwarz78}).

With respect to recent works \citep{Oguri12}, we use a larger, not homogeneous sample
and focus our attention on the determination of the cosmological constant and
the comparison with other alternative dark energy models, as the current sample size does not allow a firm determination of the rate of mass and number evolution of massive galaxies.

The paper is organized as follows. In Section~\ref{sec:method},
the basics of gravitational lensing statistics is introduced,
also allowing for number and mass evolution of the lens population.
We conduct a literature survey
for known systems, listing their basic parameters
and defining two statistical samples to perform the redshift test.
In Section~\ref{sec:result}, we introduce four cosmological models, and show the results of constraining
cosmological parameters using the Markov Chain Monte Carlo method, with and without galaxy evolution.
In Section~\ref{sec:effect}, we assess the possible presence of selection effects and
systematic biases in our galaxy sample. Finally, we present the main conclusions
and discussion in Section~\ref{sec:conclusion}.

\section{The redshift test and the sample}
\label{sec:method}

Following \citet{Ofek03}, the differential optical depth per unit redshift is
\begin{equation}
\frac{d\tau}{dz_l}=n(\Delta \theta,z_l)(1+z_l)^{3} S_\mathrm{cr} \frac{cdt}{dz_l} \, ,
\label{opt_depth}
\end{equation}
where $n(\theta,z_l)$ is the comoving number density and $S_\mathrm{cr}$ is the strong lensing cross-section.

Early-type galaxies are accurately described as singular isothermal spheres (SIS), and it is shown that radial mass distribution and ellipticity
of the lens galaxy are unimportant in altering the cosmological constraints \citep{Maoz93,Kochanek96b}. The SIS density profile is
\begin{equation}
\rho(r)=\frac{\sigma^2}{2\pi G}\frac{1}{r^2} \, ,
\end{equation}
where $\sigma$ is the velocity dispersion and $r$ is the projected distance from the center. In Section~4, we will discuss the systematic uncertainties introduced by this assumption.
The corresponding strong lensing cross-section is
\beq
\label{stat3}
S_\mathrm{cr}=16\pi^3 \left( \frac{\sigma}{c}\right)^4  \left( \frac{ D_{l}  D_{ls}}{D_{s}} \right)^2,
\eeq
where $D_{l}$, $D_{s}$, and $D_{ls}$ are the angular diameter distances between the observer and the lens,
the lens and the source
and the observer and the source, respectively.
Under a Friedman-Walker metric with null space curvature, the angular diameter
distance reads
\begin{eqnarray}
\label{inted}
D_A(z_1, z_2;\textbf{p})=\frac{c}{H_{0} (1+z_2)}\int_{z_1}^{z_2} \frac{dz'}{E(z';\textbf{p})} \, ,
\end{eqnarray}
where $H_0$ is the Hubble constant and $E(z; \textbf{p})$ is the
dimensionless expansion rate dependent on redshift $z$ and
cosmological model parameters \textbf{p}.
The two multiple images will form at an angular separation
\beq
\label{stat4}
\Delta \theta = 8 \pi \left( \frac{\sigma}{c}\right)^2  \frac{D_{ls}}{D_{s}} \, .
\eeq

We use the empirically determined velocity dispersion distribution function (VDF) of early-type galaxies: following previous
works, our sample are limited to lensing early-type galaxies.
The VDF is generally modeled by a modified Schechter function as \citep{Sheth03}
\beq
\label{stat2}
\frac{d n}{d \sigma}=
n_*\left( \frac{\sigma}{\sigma_*}\right)^\alpha \exp \left[ -\left( \frac{\sigma}{\sigma_*}\right)^\beta\right] \frac{\beta}{\Gamma (\alpha/\beta)} \frac{1}{\sigma} \, ,
\eeq
where $\alpha$, $\beta$, $n_*$, $\sigma_*$ are the faint-end slope, high-velocity cutoff, characteristic number density, and velocity dispersion, respectively.
In the following analysis we use the results of \citet{Choi07}, who analyzed data
from the SDSS Data Release 5 to derive the VDF of early-type galaxies.
The best-fit values of those VDF parameters are $n_* = 8.0{\times} 10^{-3} h^3$~Mpc$^{-3}$,
where $h$ is $H_0$ in units of 100~km~s$^{-1}$~Mpc$^{-1}$, $\sigma_*=161$~km~s$^{-1}$,
$\alpha=2.32 \pm 0.10$, and $\beta = 2.67 \pm 0.07$.

We also allow for evolution of the quantities $n_{*}$ and $\sigma_{*}$, by adopting the following parameterization:
\begin{equation}
n_{*}(z_l)= n_{*}  (1+z_l)^{P}, \sigma_{*}(z_l) = \sigma_{*} (1+z_{\rm l})^{U},
\label{s_star_z}
\end{equation}
where $P$ and $U$ are constant quantities.

In the following, we check the effect of redshift evolution by letting $P$ and $U$ be free parameters in
Section~\ref{sec:result}, instead of adopting the evolution of $(P, U) = (-0.23, -0.01)$ predicted by the semi-analytic model of \citet{Kang05,Chae07}.
Moreover, since the main goal of this paper is to constrain cosmological parameters, we first consider $P$ and $U$
as free parameters, obtain their best-fit values and probability distribution function, and marginalize them to determine constraints on the relevant cosmological parameters of interest.

Straightforward calculations lead to the optical depth per unit redshift
for a system with image separation $\Delta\theta$ and source redshift $z_{s}$,
\begin{equation}
\begin{array}{ll}
\frac{d{\tau}}{dz_l} \left( \Delta\theta,z_{s} \right) =  & \tau_{N}
             (1+z_l)^{\left[-U\alpha+P \right]} \\
        & \times (1+z_l)^{3} \frac{D_{ls}}{D_{s}} D_{l}^{2} \frac{cdt}{dz_l} \left( \frac{\Delta\theta}{\Delta\theta_{*}} \right)^{\frac{1}{2}\alpha+1} \\
        & \times \exp{[- \left( \frac{\Delta\theta}{\Delta\theta_{*}} \right)^{\frac{1}{2}\beta} (1+z_l)^{-U\beta} ]} \, ,
\end{array}
\label{dtaudz}
\end{equation}
where the normalization,
\begin{equation}
\tau_{N} = 2\pi^{2}n_{*} \frac{\beta}{\Gamma(\alpha/\beta)} \left(\frac{\sigma_{*}}{c} \right)^{2} \, ,
\label{TauNorm}
\end{equation}
and
\begin{equation}
\Delta \theta_{*} = 8\pi \left( \frac{\sigma_{*}}{c} \right)^{2} \frac{D_{\rm ls}}{D_{\rm s}} \, .
\label{theta1}
\end{equation}
For a given lens system, $d\tau /dz_{l}$ gives the
relative probability of finding the lens at different redshift:
\begin{eqnarray}
\delta p_l & = & \frac{d{\tau}}{dz_l}/\tau \nonumber\\
           & = & \frac{d{\tau}}{dz_l}/\int_{0}^{z_s}\frac{d{\tau}}{dz_l}dz_{\rm l}  \, .
\label{differential}
\end{eqnarray}
Besides the probability of lensing by a deflector at $z_{\rm l}$,  $d \tau/ d z_{\rm l}$,
the total optical depth for multiple imaging $\tau$ and the probability that a background source lensed by an SIS galaxy with image
separation $\Delta \theta$, $d \tau /d\Delta \theta$ \citep{Cao11c,Cao12a}, can also be obtained by integrating the differential probability in Equation~(\ref{dtaudz}).

Moreover, it should be noted that the image separation is taken into account as a prior in the method applied in this paper, which makes it an advantage to use almost all the known lenses \citep{Kochanek92}. We have compiled a list of 122 gravitational lenses from a variety of  sources in literature. Their basic data (lens and source redshifts both and the largest image separations) are summarized in Table~\ref{tab:data}.
As mentioned above, in order to build an homogeneous galaxy sample, we limit our analysis to
galaxies with early-type morphology.

The main source is given by the SLACS project \citep{Bolton08},
providing 59 lenses in our list. These lenses have redshifts in the range from $z_{ l}\simeq 0.05$ to 0.5, making
the lower redshift part of our overall sample,
with the lensed sources ranging from $z_{s}\simeq 0.2$ to 1.2 \citep{Bolton08}.
As a consequence of the initial spectroscopic selection method,
all the SLACS gravitational lenses
have known spectroscopic redshifts for both source and lens, giving the SLACS sample an immediate
scientific advantage over strong-lens candidate samples selected from imaging data.
However, it is known that the SLACS sample
is biased towards moderately large-separation lenses ($\Delta \theta > 2''$; \citep{Arneson12}),
leading to biased estimates in the redshift test \citep{CN07}.

A majority of the sample was observed as part of
the CASTLES program \footnote{http://cfa-www.harvard.edu/castles/}, but it also contains
gravitational lenses found in the COSMOS survey\footnote{http://cosmos.astro.caltech.edu/}\citep{Faure08}
and the Extended Growth Strip (EGS; \citep{Moustakas07}),
including six additional
COSMOS and EGS systems discovered recently:
COSMOS5921+0638, COSMOS0056+1226, COSMOS0245+1430, "Cross", and "Dewdrop".
Finally, we also include five early-type gravitational lenses from
Lenses Structure and Dynamics survey (LSD, \citep{Koopmans02,Koopmans03,Treu04}), spanning the redshift range
$0.48< z_{l}<1.00$ (Q0047-2808, HST15433+5352, MG2016, CY2201-3201 and CFRS03.1077).

Moreover, lenses dominated by a group or cluster potential will affect the constraint results
\citep{Keeton04,Oguri05,Faure11} and previous versions of the lens redshift test have certainly gone to some effort to exclude these systems.
Therefore, systems known to be strongly affected by the presence of a group (e.g., B1359+154) or galaxy
cluster (e.g., Q0957+561, SDSS1004+4112, and B2108+213) have been excluded from our sample.
Moreover, we adopt the image separation criterion of $4''$ to remove lenses that are influenced by complex environments such as clusters \citep{Ofek03}. On this basis we have removed Q0047-2828, and RXJ0921+4529 from the final sample. However, it should be noted that this technique is not biased by this criterion, because the redshift of a lens that has a large separation is very low compared with the source redshift\citep{Ofek03}.

The redshift distribution test requires a statistically complete
and well characterized sample. As our list includes galaxies from a variety of surveys,
using very different observational strategies and discovery spaces
(the SDSS spectroscopic sample, Hubble Space Telescope (HST) field sky survey, etc.),
it is mandatory to verify the completeness of our final sample of gravitational lenses
and its usability for the redshift test. In our analysis, we will use the two following samples (see Table~\ref{subsample}).

Sample A. Sixty-three lenses from the above list, excluding the whole SLACS sample.
This sample is extracted by the same parent population as the primary sample investigated in \citet{CN07}.

Sample B. Seventy-one lenses with image separation larger than $2''$.
This choice is motivated by the fact that
the SLACS sample is biased toward moderately large-separation lenses and
large velocity dispersions
and is less than 50\% complete below $\Delta \theta = 2''$ \citep{Arneson12}.

We will discuss the samples selection functions and the impact on our results in Section~\ref{sec:effect}.

\begin{table*}
\centering
\caption{\label{tab:data}Summary of the Properties of the Strongly Lensed Systems, with the SLACS Lenses Written in Bold. References:
$1$ - \citet{Bolton08};
$2$ - \citet{Treu06};
$3$ - \citet{Newton11};
$4$ - \citet{Ruff11};
$5$ - \citet{Koopmans02};
$6$ - \citet{Koopmans03};
$7$ - \citet{Treu04};
$8$ - \citet{Kochanek00};
$9$ -\citet{Winn02b};
$10$ -\citet{Gregg02};
$11$ -\citet{Hall02};
$12$ -\citet{ODea92};
$13$ -\citet{Wiklind95};
$14$ -\citet{Cohen02b};
$15$ -\citet{Crampton02};
$16$ -\citet{Falco97};
$17$ -\citet{Tonry99};
$18$ -\citet{Wisotzki02};
$19$ -\citet{Gregg00};
$20$ -\citet{Fassnacht98};
$21$ -\citet{Hagen00};
$22$ -\citet{Lubin00};
$23$ -\citet{Kneib00};
$24$ -\citet{Munoz01};
$25$ -\citet{Schechter98};
$26$ -\citet{Lehar00};
$27$ -\citet{Lacy02};
$28$ -\citet{Hewett94};
$29$ -\citet{Surdej97};
$30$ -\citet{Eisenhardt96};
$31$ -\citet{Lidman00};
$32$ -\citet{Tonry98};
$33$ -\citet{Myers99};
$34$ -\citet{Siemiginowska98};
$35$ -\citet{Ratnatunga99};
$36$ -\citet{Burud02b};
$37$ -\citet{Lehar93};
$38$ -\citet{Fassnacht96};
$39$ -\citet{Winn02a};
$40$ -\citet{Morgan01};
$41$ -\citet{Langston89};
$42$ -\citet{Wiklind96};
$43$ -\citet{Winn00};
$44$ -\citet{Sykes98};
$45$ -\citet{Fassnacht99};
$46$ -\citet{Burud02a};
$47$ -\citet{Huchra85};
$48$ -\citet{Inada05};
$49$ -\citet{Biggs03};
$50$ -\citet{Johnston03};
$51$ -\citet{Blackburne08};
$52$ -\citet{Sluse03};
$53$ -\citet{Eigenbrod06b};
$54$ -\citet{Inada08};
$55$ -\citet{Morgan04};
$56$ -\citet{Ofek06};
$57$ -\citet{Faure08};
$58$ -\citet{Anguita09};
$59$ -\citet{Lagattuta10};
$60$ -\citet{Moustakas07}.
}
{\scriptsize
\begin{tabular}{llllllllllllllll}

\hline
 Lens Name & $z_{s}$ & $z_{l}$ & $\Delta\theta$($''$) & Ref  &  Lens Name & $z_{s}$ & $z_{l}$ & $\Delta\theta$($''$) & Ref\\
\hline
\textbf{J0029-0055}	&  0.9313	&  0.227	&  1.92  & 1    &J221929-001743       &1.0232    &0.2888     &1.472   & 4 \\
\textbf{J0037-0942}	&  0.6322	&  0.1955	&  2.943 & 1,2  &J022511-045433       &1.1988    &0.2380     &3.54    & 4 \\
\textbf{J0044+0113}	&0.1965	    &0.1196	    &1.58    & 1,3  &J022610-042011   &1.232  &0.4943      &2.306   & 4 \\
\textbf{J0109+1500}	&0.5248	    &0.2939	    &1.38    & 1    &MG2016	       &3.263  &1.004  &3.12   & 5\\
\textbf{J0216-0813}	&0.5235	    &0.3317	    &2.303   & 1,2,3  &HST15433+5352    &2.092 &0.497	&1.18   & 5,6,7\\
\textbf{J0252+0039}	&0.9818	    &0.2803	    &2.08    & 1    &CY2201-3201	   &3.900  &0.320  &0.830   & 5,6,7\\
\textbf{J0330-0020}	&1.0709	    &0.3507	    &2.2     & 1,3  &CFRS03.1077	   &2.941  &0.938  &2.48    & 5,6,7\\
\textbf{J0405-0455} &0.8098	    &0.0753	    &1.6     & 1    &Q0142-100        & $2.72$  & $0.49$  & $2.231$   & 8 \\
\textbf{J0728+3835}	&0.6877	    &0.2058	    &2.5     & 1    &PMNJ0134-0931    & $2.216$ & $0.76451$ &$0.7$    & 9,10,11 \\
\textbf{J0737+3216}	 &0.5812	&0.3223	    &2.065   & 2,3  &B0218+357         & $0.944$ & $0.685$  &$0.33$    & 12,13,14\\
\textbf{J0822+2652}	 &0.5941	&0.2414	    &2.34    & 1    &CFRS03.1077       & $2.941$ & $0.938$  &$2.1$    & 15    \\
\textbf{J0841+3824}	 &0.6567	&0.1159	    &2.82    & 1    &MG0414+0534       & $2.64$  & $0.9584$ &$2.12$   & 16,17 \\
\textbf{J0912+0029}	 &0.3239	&0.1642	    &3.23    & 1,2  &HE0435-1223       & $1.689$ & $0.4$    &$2.6$   & 18  \\
\textbf{J0935-0003}	 &0.467	    &0.3475	    &1.74    & 1,3  &HE0512-3329       & $1.565$ &$0.9313$   &$0.644$ & 19  \\
\textbf{J0936+0913}	 &0.588	    &0.1897	    &2.18    & 1    &B0712+472         & $1.339$ & $0.406$ & $1.28$ & 20  \\
\textbf{J0946+1006}	 &0.6085	&0.2219	    &2.76    & 1    &MG0751+2716       & $3.200$ & $0.3502$ & $0.7$ & 17  \\
\textbf{J0955+0101}	 &0.3159	&0.1109	    &1.82    & 1    &HS0818+1227       & $3.115$ & $0.39$  & $2.55$   & 21  \\
\textbf{J0956+5100}	 &0.4699	&0.2405	    &2.642   & 1,2  &SBS0909+523        & $1.377$ & $0.830$ & $1.10$   & 22  \\
\textbf{J0959+4416}	 &0.5315	&0.2369	    &1.92    & 1,2  &RXJ0911+0551      & $2.80$  & $0.77$ & $3.25$   & 23 \\
\textbf{J0959+0410}	 &0.535	    &0.126	    &1.995   & 1    &FBQ0951+2635       & $1.24$  & $0.25$ & $1.10$  & 25  \\
\textbf{J1016+3859}	 &0.4394	&0.1679	    &2.18    & 1    &BRI0952-0115      & $4.50$  & $0.41$& $0.99$ & 26  \\
\textbf{J1020+1122}	 &0.553	    &0.2822	    &2.4     & 1    &J100424.9+122922  & $2.65$  & $0.95$& $1.54$ & 27 \\
\textbf{J1023+4230}	 &0.696	    &0.1912	    &2.82    & 1    &LBQS1009-0252     & $2.74$  & $0.88$& $1.53$ & 28  \\
\textbf{J1029+0420}	 &0.6154	&0.1045	    &2.02    & 1    &Q1017-207         & $2.545$ & $1.085$ & $0.849$ & 29 \\
\textbf{J1032+5322}	 &0.329	    &0.1334	    &2.06    & 1    &FSC10214+4724     & $2.286$ & $0.914$ & $1.59$  & 30  \\
\textbf{J1103+5322}	 &0.7353	&0.1582	    &2.04    & 1,3  &B1030+071          & $1.535$ & $0.599$ & $1.56$   & 20  \\
\textbf{J1106+5228}	 &0.4069	&0.0955	    &2.46    & 1,3  &HE1104-1805        & $2.32$  & $0.729$ & $3.19$     & 31  \\
\textbf{J1112+0826}	 &0.6295	&0.273	    &2.98    & 1,3  &PG1115+080        & $1.72$  & $0.311$&$2.42$ & 32  \\
\textbf{J1134+6027}	 &0.4742	&0.1528	    &2.2     & 1    &B1152+200          & $1.019$ & $0.439$  & $1.56$    & 33\\
\textbf{J1142+1001}	 &0.5039	&0.2218	    &1.96    & 1,3  &Q1208+101          & $3.80$  & $1.1349$ & $0.47$  & 34 \\
\textbf{J1143-0144}  &0.4019	&0.106	    &3.36    & 1    &HST14113+5211      & $2.811$ & $0.465$ & $2.26$   & 22  \\
\textbf{J1153+4612}	 &0.8751	&0.1797	    &2.1     & 1    &HST14176+5226      & $3.40$  & $0.81$  & $3.25$   & 35  \\
\textbf{J1204+0358}   &0.6307	&0.1644	    &2.62    & 1,3  &B1422+231         & $3.62$  & $0.339$ &$1.28$   & 32 \\
\textbf{J1205+4910}	 &0.4808	&0.215	    &2.44    & 1    &SBS1520+530        & $1.855$ & $0.717$ &$1.568$  & 8,36\\
\textbf{J1213+6708}	 &0.6402	&0.1229	    &2.84    & 1,3  &MG1549+3047        & $1.17$ & $0.11$ &$2.3$   & 37 \\
\textbf{J1218+0830}	 &0.7172	&0.135	    &2.9     & 1,3  &B1600+434          & $1.589$ & $0.4144$ & $1.38$   & 20  \\
\textbf{J1250+0523}	 &0.7953	&0.2318	    &2.26    & 1,2  &B1608+656          & $1.394$ & $0.630$ & $2.27$    & 38 \\
\textbf{J1330-0148}   &0.7115   &0.0808     &1.706   & 1,2  &PMNJ1632-0033      & $3.424$ & $1.0$ & $1.47$  & 39   \\
\textbf{J1402+6321}   &0.4814	&0.2046	    &2.775   & 1,2  &FBQ1633+3134       & $1.52$  & $0.684$ & $0.66$ & 40  \\
\textbf{J1403+0006}    &0.473	&0.1888	    &1.66    & 1    &MG1654+1346        & $1.74$  & $0.254$ & $2.1$   & 41  \\
\textbf{J1416+5136}    &0.8111	&0.2987	    &2.74    & 1,3  &PKS1830-211         & $2.51$  & $0.886$ & $0.99$    & 42  \\
\textbf{J1420+6019}   &0.5351	&0.0629	    &2.097   & 1,2  &PMNJ1838-3427       & $2.78$  & $0.31$ & $1.00$     & 43 \\
\textbf{J1430+4105}   &0.5753	&0.285	    &3.04    & 1    &B1933+507          & $2.63$  & $0.755$ & $1.00$   & 44 \\
\textbf{J1432+6317}   &0.6643	&0.123	    &2.52    & 1,3  &B2045+265          & $1.28$  & $0.8673$ & $2.2$    & 45 \\
\textbf{J1436-0000}   &0.8049	&0.2852	    &2.24    & 1,3  &HE2149-2745        & $2.03$  & $0.50$ &$1.69$  & 8, 46 \\
\textbf{J1443+0304}    &0.4187	&0.1338	    &1.62    & 1,3   &Q2237+0305         & $1.695$ & $0.0394$ & $1.82$   & 47 \\
\textbf{J1451-0239}	 &0.5203	&0.1254	    &2.08    & 1,3  &SDSS0246-0825      &1.68  &0.724         &1.2 & 48 \\
\textbf{J1525+3327}	 &0.7173	&0.3583	    &2.62    & 1,3  &B0850+054          &3.93   &0.59         &0.68 & 49 \\
\textbf{J1531-0105}	 &0.7439	&0.1596	    &3.42    & 1,3  &SDSS0903+5028      &3.605 &0.388         &3.0  & 50 \\
\textbf{J1538+5817}	  &0.5312	&0.1428	    &2       & 1,3  &HE1113-0641        &1.235   &0.75     &0.88 & 51 \\
\textbf{J1621+3931}	 &0.6021	&0.2449	    &2.58    & 1,3  &Q1131-1231         &0.658   &0.295     &3.8  & 52 \\
\textbf{J1627-0053}	  &0.5241	&0.2076	    &2.42    & 1,2  &SDSS1138+0314      &2.44  &0.45        &1.34  & 53, 54 \\
\textbf{J1630+4520}	  &0.7933	&0.2479	    &3.618   & 1,2  &SDSS1155+6346      &2.89    &0.176      &1.96  & 55 \\
\textbf{J1636+4707}	  &0.6745	&0.2282	    &2.18    & 1    &SDSS1226-0006      &1.12  &0.52      &1.26  & 53, 54 \\
\textbf{J2238-0754}	 &0.7126	&0.1371	    &2.54    & 1,3  &WFI2033-4723       &1.66  &0.66     &2.34  & 53, 55 \\
\textbf{J2300+0022}	 &0.4635	&0.2285	    &2.494   & 1,2  &HE0047-1756     &1.66  &0.41         &1.54 & 56 \\
\textbf{J2303+1422}	 &0.517	    &0.1553	    &3.278   & 1,2  &COSMOS5921+0638   &3.15  &0.551     &1.6 & 57, 58 \\
\textbf{J2321-0939}	 &0.5324	&0.0819	    &3.2     & 1,2  &COSMOS0056+1226   &0.81 &0.361      &2.4 & 57, 59\\
\textbf{J2341+0000}  &0.807	    &0.186	    &2.88    & 1,3  &COSMOS0245+1430  &0.779   &0.417     &3.08 & 57,59\\
J021737-051329       &1.847     &0.6458     &2.536   & 4    &"Cross"          &3.40   &0.810      &2.44 & 60 \\
J141137+565119       &1.420     &0.3218     &1.848   & 4    &"Dewdrop"        &0.982   &0.580     &1.52 & 60  \\
\hline
\end{tabular}}
\end{table*}

\begin{table}
\caption{\label{subsample} Summary of the Subsamples Used in the Analysis of This Paper.}
\begin{center}
\begin{tabular}{c|l}\hline\hline
 Subsample & \hspace{4mm}Definition\hspace{4mm}\\ \hline

 Sample A & 63 lenses from Table \ref{tab:data} excluding the SLACS sample \\
 Sample B & 71 lenses with image separation larger than $2''$ \\
 Sample C & 51 lenses with image separation not larger than $2''$\\
 SLACS    & 59 lenses from the whole SLACS sample \\
 Full sample & 122 lenses from Table \ref{tab:data} \\
 \hline\hline
\end{tabular}
\end{center}
\end{table}

\section{Statistical analysis}
\label{sec:result}

Our statistical analysis is based on the maximum likelihood technique.
For a sample of $N_{\rm{L}}$ multiply imaged sources, the likelihood
$\mathcal{L}$ of the observed lens redshift
given the statistical lensing model is defined by
\begin{equation}
\ln \mathcal{L} = \sum_{l=1}^{N_{\rm{L}}}\ln \delta p_l(\bf{p}),
\label{lnL}
\end{equation}
where $\delta p_l(\bf{p})$ is the particular differential probability given
by Equation(\ref{differential}) normalized to one, and $\bf{p}$ are the cosmological model parameters
(e.g., $\Omega_{\Lambda}$, $\Omega_{m}$), for the $l$th multiply imaged source.
Accordingly, the $\chi^2$ is defined as follows:
\begin{equation}
\chi^2 = -2 \ln \mathcal{L} \, .
\label{chi}
\end{equation}
and he best-fit model parameters are determined by minimizing $\chi^2$. Our analysis is based on the publicly available package COSMOMC \citep{Lewis02}.

\vspace{0.2cm}

We consider four different cosmological models
to be tested with the observed lens redshift distribution:
the $\Lambda$CDM model and three phenomenological models in which the vacuum energy is described
as a dynamical quantity:
the so-called XCDM model with the EoS $w=p/\rho$ a free parameter,
the DGP model arising from the brane world theory, and the RDE models.
These models are
motivated by the well known fine-tuning and coincidence problems of the standard $\Lambda$CDM model.

We note that the previous precision cosmological observational
data have hinted that both the RDE and dark energy with EoS $w<-1$ may have dubious stability problems \citep{Feng09, Amani11},
and the DGP model has already been ruled out observationally \citep{Fang08,Durrer10,Maartens10},
so it is indicated that these are just supposed to be a representative set, instead of viable candidates for dark energy.

It is well known that the $\chi^2$-statistics alone is not sufficient to provide an
effective way to make a comparison between different models. In this paper, we use the BIC as a model selection criterion \citep{Schwarz78}. The BIC is defined by
\begin{equation}
{\rm BIC}=-2\ln{\cal L}_{max}+k\ln N \, ,
\end{equation}
where ${\cal L}_{\rm max}$, $k$, $N$ are the maximum likelihood, the number of parameters, and the
number of data points, respectively. Under this selection criterion, a positive evidence against the model with the higher BIC is defined by a difference $\Delta{\rm BIC}=2$ and a strong evidence is defined by $\Delta{\rm BIC}=6$.

A spatially flat universe is assumed throughout the paper, which is strongly supported a combined 5-years Wilkinson Microwave
Anisotropy Probe (WMAP5), BAOs, and SN data \citep{Hinshaw09}.
As mentioned above, we also consider the case of an evolving population of lensing galaxies
in order to assess the accuracy of our results.
We consider simultaneous constraints on the galaxy evolution and cosmological
parameters. In order to derive the probability distribution function for the cosmological parameters of interest,
we marginalize $P$ and $U$ and perform fits of different cosmological scenarios on
both Samples A and B.
Results are shown in Figure~\ref{L} - \ref{RDE} and summarized in
Table~\ref{tab:result}.

\subsection{The standard cosmological model ($\Lambda$CDM)}

\begin{figure}
   \centering
   \includegraphics[width=0.5\textwidth]{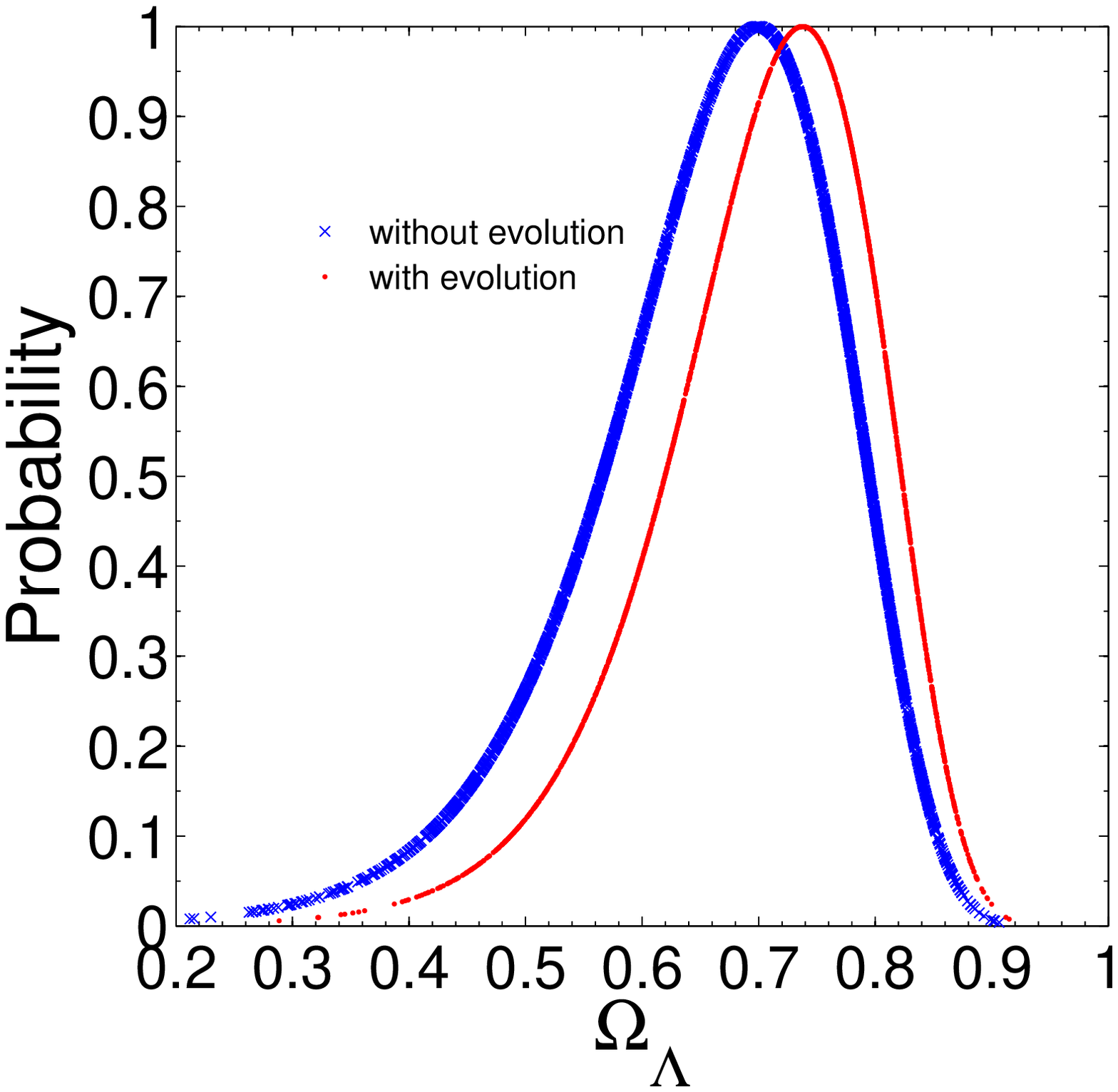}
   \includegraphics[width=0.5\textwidth]{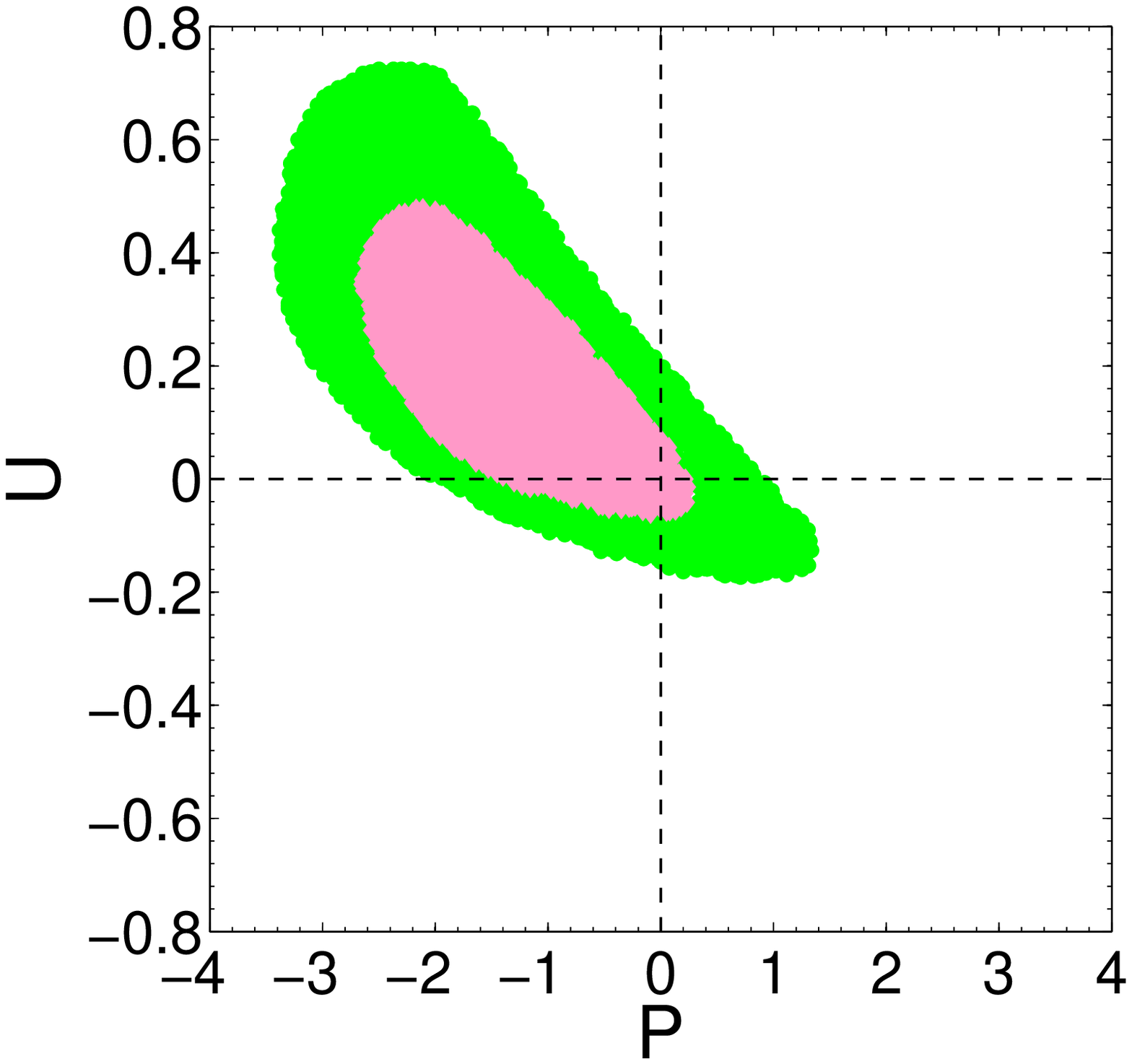}
\caption{Simultaneous constraints on the cosmological constant and VDF evolutions in the flat $\Lambda$CDM model
obtained by using Sample A. Upper panel: likelihood distributions as a function of $\Omega_\Lambda$ with and without redshift evolution.
The red dotted line is obtained after marginalizing over $P$ and $U$.
Lower panel: constraints on $P$ and $U$ after marginalizing over $\Omega_\Lambda$.
Dashed lines in the lower panel represent the case without redshift evolution ($P=U=0$). }
\label{L}
\end{figure}

\begin{figure}
   \centering
   \includegraphics[width=0.5\textwidth]{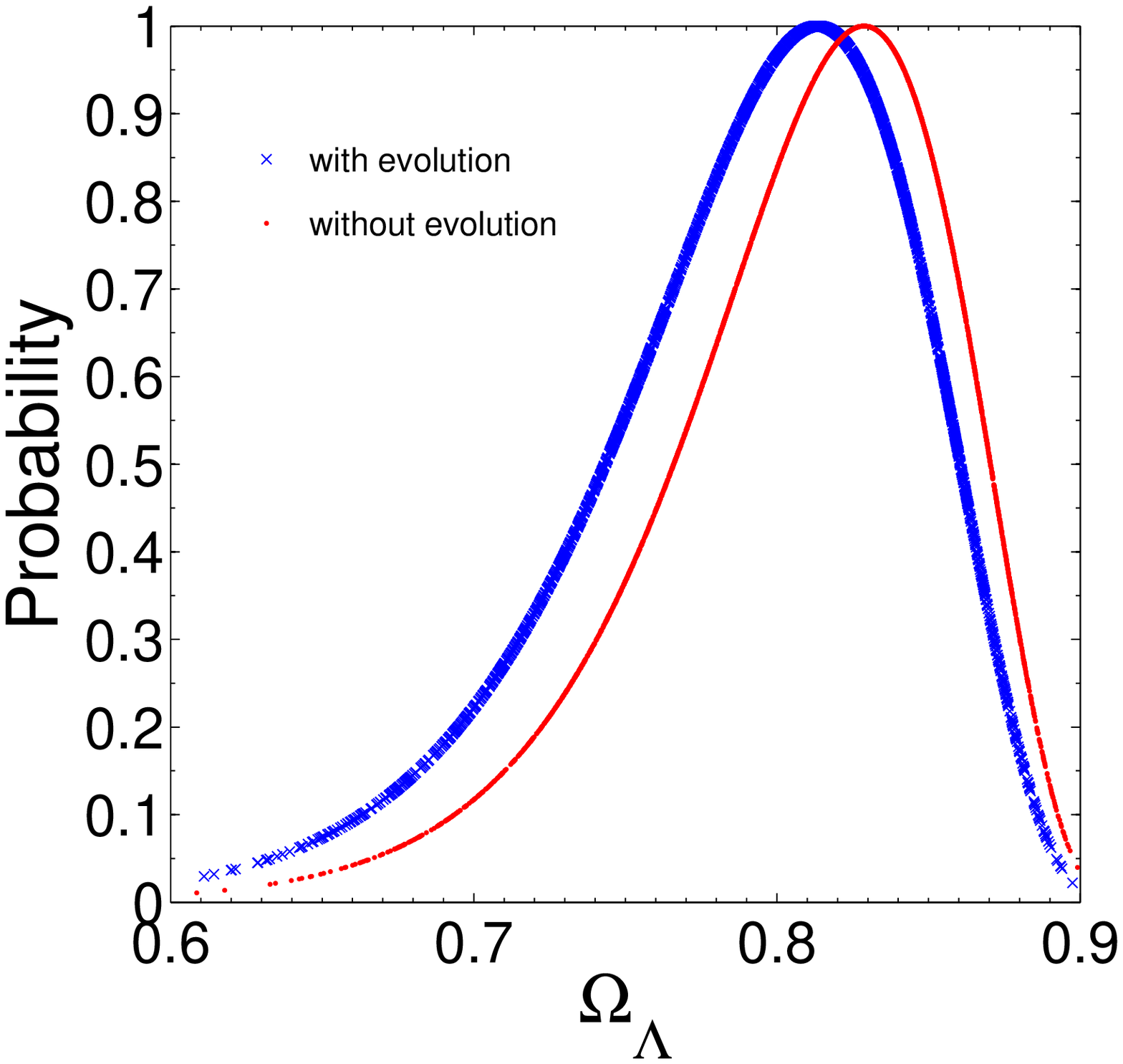}
   \includegraphics[width=0.5\textwidth]{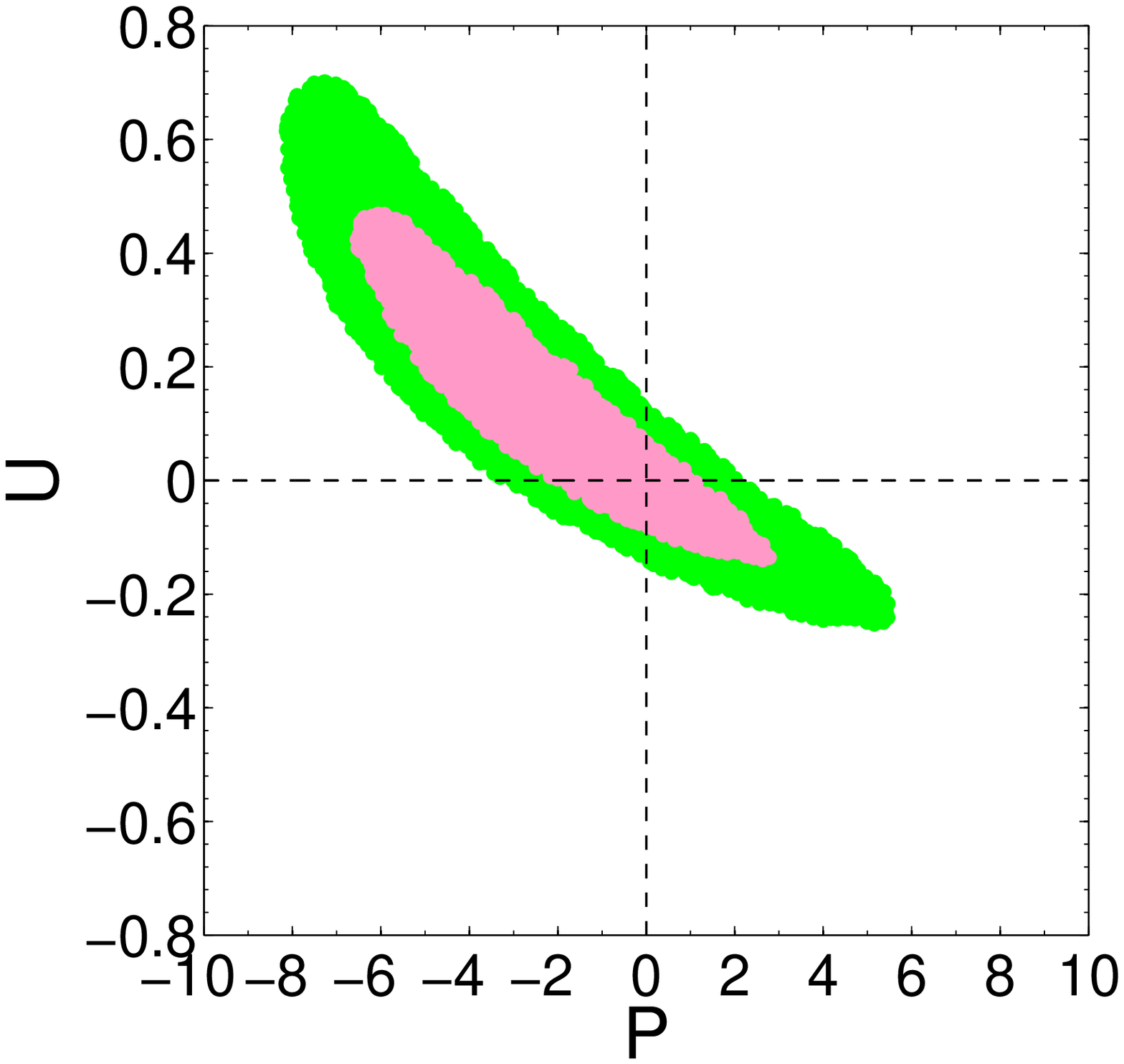}
\caption{As in Figure~\ref{L}, using Sample B (i.e., lensing systems with image separation larger than $2''$).}
\label{L2}
\end{figure}


In the simplest scenario, the
dark energy is a cosmological constant, $\Lambda$, i.e., a
component with constant EoS $w=p/\rho=-1$.
If spatial flatness of the Friedmann-Robertson-Walker metric is assumed,
the Hubble parameter according to the Friedmann equation is
\begin{equation}
H^2 = H_0^2 [\Omega_m(1+z)^3+\Omega_\Lambda] \, ,
\end{equation}
where $\Omega_m$ and $\Omega_\Lambda$ represent the density parameters of
matter (both baryonic and non-baryonic components) and cosmological constant, respectively.
As $\Omega=\Omega_m+\Omega_\Lambda=1$, this model
has only one independent parameter.

We consider constraints obtained both samples defined above.
While considering Sample A,
the likelihood is maximized, ${\cal L}={\cal L_\mathrm{max}}$, for $\Omega_\Lambda=0.70\pm 0.09$
with no redshift evolution and $\Omega_\Lambda= 0.73\pm 0.09$ with redshift evolution,
see Figure~\ref{L}. Hereafter, uncertainties denote the statistical 68.3\% confidence limit for one parameter,
determined by ${\cal L}/{\cal L_\mathrm{max}}=\exp (-1/2)$.
Data are consistent with the no-redshift-evolution case ($P=0, \, U=0$)
at $1\sigma$. Specifically, the measured values of the two parameters are
$P=-1.2\pm 1.4$ and $U=0.22_{-0.27}^{+0.26}$.

When using the Sample B, we find, in the no-evolution scenario, $\Omega_\Lambda = 0.81 \pm 0.05$,
consistent with the result from Sample A.
When allowing for galaxy evolution, we find $ \Omega_\Lambda=0.83 \pm 0.05$,
$P=-1.9 \pm 4.6$, and $U=0.16 \pm 0.30$.

In both cases, our findings are very close to the ones obtained
from the ESSENCE supernova survey data, $\Omega_\Lambda=0.73\pm 0.04$ in the flat case \citep{Davis07},
and from the combined WMAP 5-year, BAO, and SN Union data \citep{Komatsu09}
with the best-fit parameter $\Omega_\Lambda=0.726 \pm 0.015$.
Moreover, both samples used here exclude with large confidence (4$\sigma$ level)
the null hypothesis of a vanishing  $\Omega_\Lambda$, as
obtained also by \citet{Oguri12} in their statistical analysis on the SQLS data,
providing independent evidence of the accelerated expansion.

While detailed analysis on the constraints of the redshift distribution test
on the hierarchical models of galaxy evolution is beyond the scope of this work,
we notice that these results obtained with both samples are in broad agreement
with previous studies \citep{CN07,Oguri12},
in which no strong evidence for any evolution of the parameters $U$ and $P$ was found.
Moreover, we find that the degenerate direction in the evolution parameters  corresponds to a constant lensing probability, as already been noted by \citet{Oguri12}.

Previous studies on lensing statistics \citep{Cha03,Ofek03}
considering a not-evolving velocity dispersion function have obtained consistent results with the galaxy number counts \citep{Im02}.
\citet{Mit05,CN07} assumed a non-evolving shape for the VDF
and obtained results consistent with earlier results. To sum up, all
previous results on redshift evolution from strong lensing statistics
appear to be in agreement with no-evolution of early-type
galaxies. As we only considered early-type lensing galaxies,
our results further confirm this conclusion.

Substantial evolution of galaxy number (i.e, $P\simeq -1 $) and mass
(i.e., $U \simeq 0.25$) are supported by studies of early-type galaxies from $z=1$ to 0 \citep{Faber07,Brown07,Oguri12}
This conclusion is consistent with our results in Section~\ref{sec:result}.

\subsection{Constraints on selected dark energy models}

Now we focus on the constraints obtained on selected
dark energy models. Here, we only use Sample A, as results obtained with Sample B are
coherent at $1\sigma$ level.

\subsubsection{Dark energy with constant equation of state (XCDM)}
While deviating from the simple case $w=-1$, the EoS of dark energy $w<-1/3$ can also give birth to an
accelerated university expansion. In a zero-curvature universe, the Hubble parameter reads:
\begin{equation}
H^2 = H_0^2 [\Omega_m(1+z)^3+\Omega_x(1+z)^{3(1+w)}] \, .
\end{equation}
Obviously, when flatness is assumed
($\Omega_m+\Omega_\Lambda=1$), it is a two-parameter cosmological model,
${\textbf{p}}=\{\Omega_x,~w\}$.

\begin{figure}
\centering
\includegraphics[width=8.9cm]{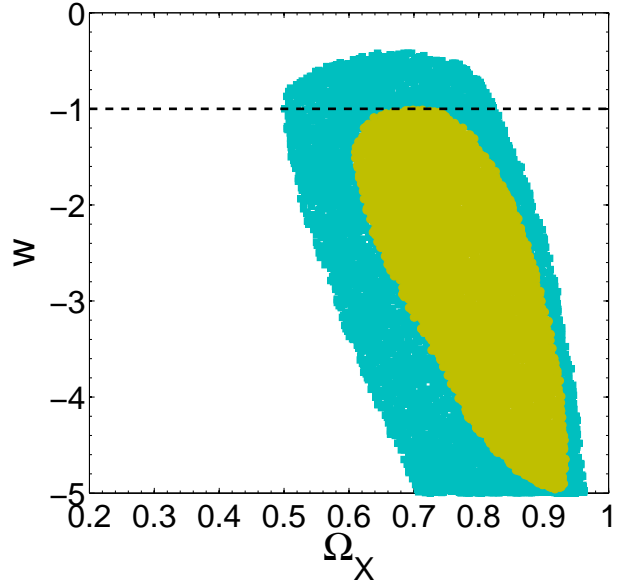}
\caption{ Likelihood contours for the flat XCDM model at 68.3\% and 95.4\% CL in the $\Omega_x-w$
plane obtained by using Sample A with no redshift evolution. The lower limit on $w$ is probably an artifact of the prior $w>-5$} and the horizontal dotted line indicates a cosmological constant with $w=-1$.
\label{w1}
\end{figure}

\begin{figure}
\centering
\includegraphics[width=0.5\textwidth]{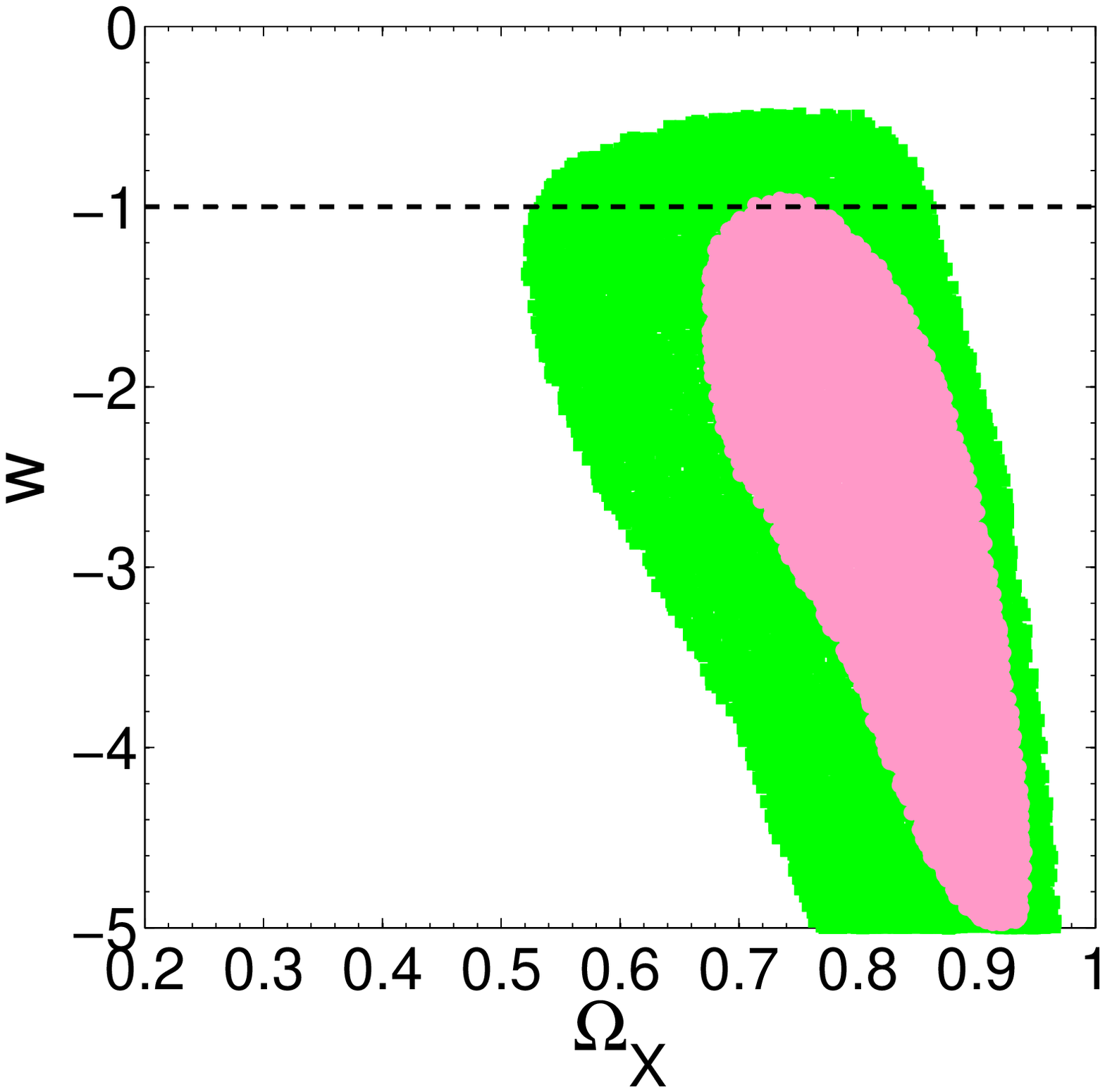}
\includegraphics[width=8.9cm]{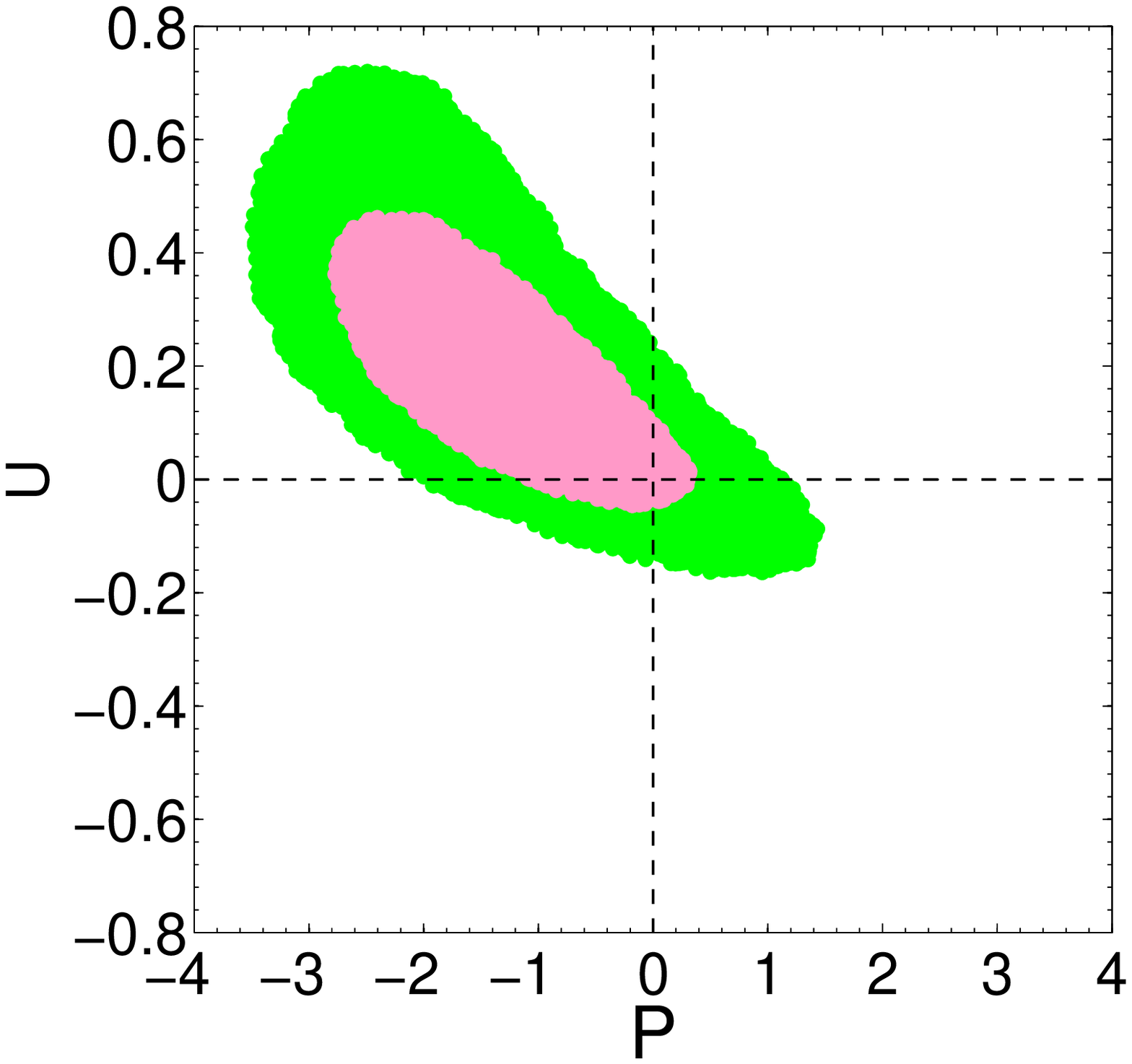}
\caption{ As in Figure~\ref{L}, but for the flat XCDM model obtained by using Sample A with redshift evolution. The lower limit on $w$ is probably an artifact of the prior $w>-5$} and the horizontal dotted line indicates a cosmological constant with $w=-1$.
\label{w2}
\end{figure}

The best-fit values of the parameters are: $\Omega_x= 0.77\pm0.17$; $w= -2.3_{-2.7}^{+1.3}$
with no redshift evolution and $\Omega_x= 0.79\pm0.13$; $w= -2.1_{-2.8}^{+1.1}$ with redshift evolution,
see Figure~\ref{w1} and Figure~\ref{w2} for the confidence limits in the $\Omega_x-w$ plane.
However, we note that the lower limits on the parameter $w$ are probably an artifact of the prior $w>-5$,
which may be tested and constrained with future larger lens sample.
Also in this cosmological scenario, the lens redshift data are consistent with no redshift evolution:
when marginalizing over $\Omega_x$, we find
$P=-1.4 \pm 1.4$ and $U=0.20 \pm 0.28$.
The Einstein's cosmological constant ($w=-1$)
is still consistent within 1$\sigma$.
Meanwhile, compared to the
cosmological constant model, this flat cosmology with constant EoS
dark energy provides a lower $\chi_{min}^2$, but a higher BIC:
$\Delta{\rm BIC}=1.72$ with no redshift evolution and $\Delta{\rm BIC}=1.55$ with redshift evolution.
However, we note that comparing with the
cosmological constant model, the two-parameter XCDM model
performs relatively well under the information criterion test.

\subsubsection{Dvali-Gabadadze-Porrati model (DGP)}

In a brane world theory, the accelerated expansion of the universe can be explained by the gravity
leaking out into the bulk at large scales with a crossover scale \citep{Dvali00}.
The Friedmann equation is
\begin{equation}
3M_{\rm Pl}^2\left(H^2-{H\over r_c}\right)=\rho_m(1+z)^3 \, ,
\end{equation}
where $M_{\rm Pl}$ is the Planck mass and $r_c=(H_0(1-\Omega_m))^{-1}$ is the crossover scale at which the induced 4-dimensional Ricci
scalar dominates. In this model, the Hubble parameter expresses as
\begin{equation}
H^2 = H_0^2 (\sqrt{\Omega_{m}(1+z)^3+\Omega_{r_c}}+\sqrt{\Omega_{r_c}})^2 \, ,
\end{equation}
and the DGP model has one one free parameter $\textbf{p}=\{\Omega_m\}$ in a flat universe.

\begin{figure}
   \centering
   \includegraphics[width=9.0cm]{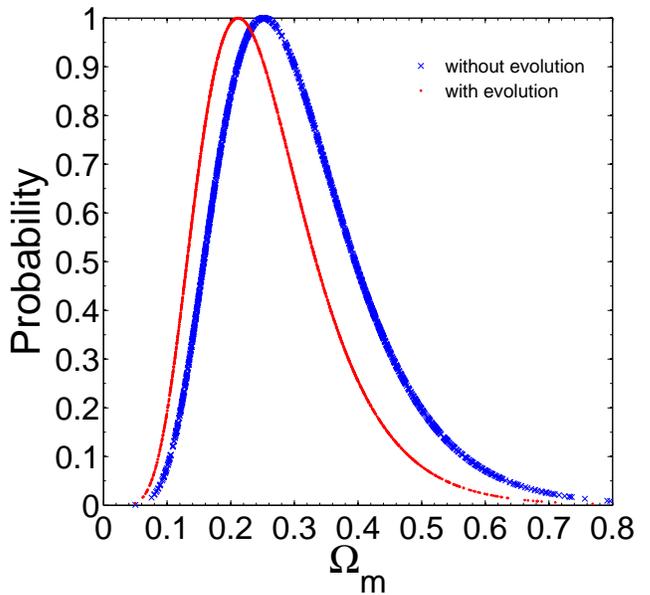}
\caption{ Normalized likelihood as a function of $\Omega_m$ for the DGP model obtained by using Sample A with and without redshift evolution.}
\label{DGP}
\end{figure}

For the DGP model, the best-fit parameters are $\Omega_m= 0.25_{-0.09}^{+0.11}$ with no redshift evolution and $\Omega_m= 0.22_{-0.09}^{+0.10}$ with redshift evolution (See Figure~\ref{DGP}). We find that the DGP model is somehow worse than the $\Lambda$CDM model with this observational data. While its $\chi^2_{\rm min}$ is larger than that of the $\Lambda$CDM model by
about 1.2, it gives $\Delta{\rm BIC}=1.20$ with no redshift evolution and $\Delta{\rm BIC}=1.19$ with redshift evolution.

\subsubsection{Ricci dark energy (RDE) model}

Following \citet{Gao09,Li10}, the average radius of the Ricci scalar curvature $|{\cal R}|^{-1/2}$
might provide an infrared cutoff length scale. Accordingly, the DE energy density is
\begin{equation}
\rho_{de}=3\beta ^2(\dot{H}+2H^2) \, ,
\label{rde}
\end{equation}
and the Hubble parameter reads
\begin{equation}
H^2 = H_0^2 [\frac{
2\Omega_{m}}{2-\beta}(1+z)^{3}+(1-{2\Omega_{m}\over
2 -\beta})(1+z)^{(4-{2\over\beta})}] \,
\end{equation}
where $\beta$ is a positive constant to be determined. This is a two-parameter model with  $\textbf{p}=\{\Omega_m,~\beta\}$.

\begin{figure}
   \centering
   \includegraphics[width=8.9cm]{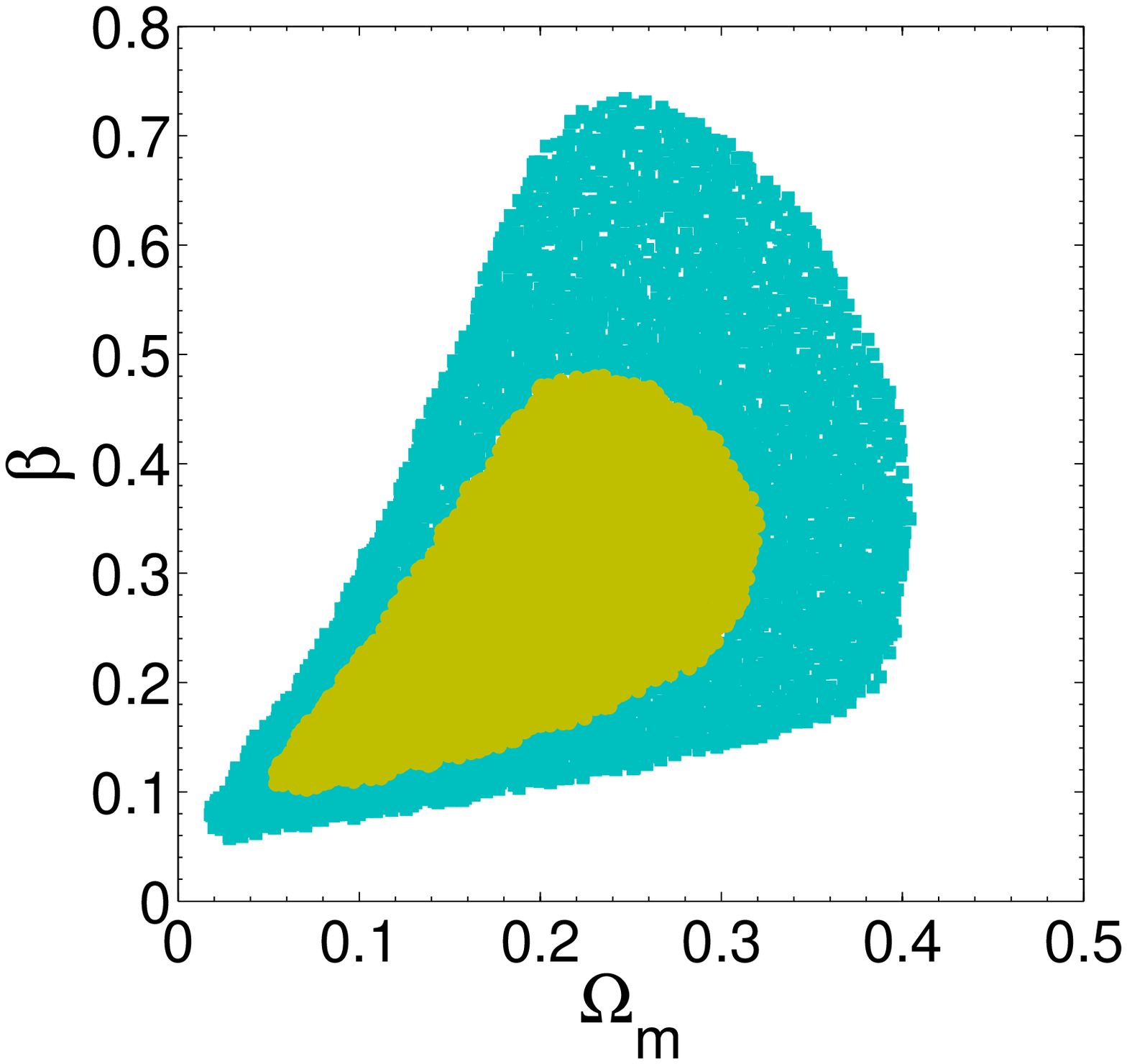}
   \includegraphics[width=8.9cm]{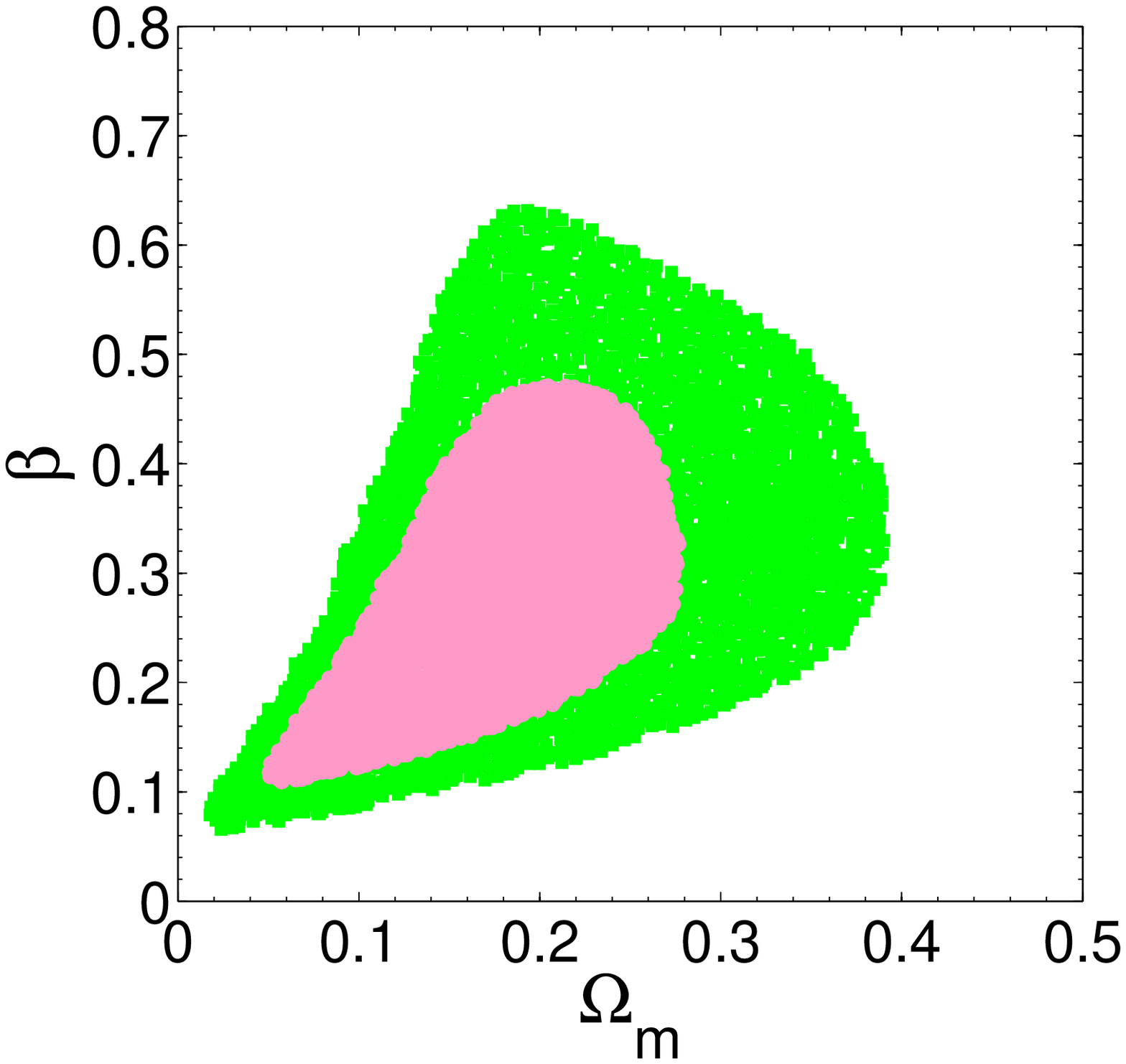}
\caption{ Likelihood contours for the RDE model at 68.3\% and 95.4\% CL obtained by using Sample A with no redshift evolution (upper) and with redshift evolution (lower).}
\label{RDE}
\end{figure}

For the RDE model, the best-fit values of the parameters are
$\Omega_m=0.22^{+0.10}_{-0.11} \, ; \,   \beta =0.29\pm 0.19$ with no redshift evolution,
and $\Omega_m=0.18^{+0.11}_{-0.12}; \beta =0.28\pm 0.18$ with redshift evolution.
In Figure~\ref{RDE}, we plot the likelihood contours in the
$\Omega_m-\beta$ plane. Given the largest information criterion result:
$\Delta{\rm BIC}=1.78$ with no redshift evolution, and $\Delta{\rm BIC}=1.62$ with redshift evolution, the RDE model
performs the worst in all the cosmological models considered in this paper.

\section{Biases and possible systematic effects}
\label{sec:effect}

Thus far we have considered only statistical errors.
Indeed, cosmological tests based on strong lensing have been somehow controversial since \citet{Kochanek96a},
in particular for the possible biases associated with sample selection \citep{CN07}.
In this section,  we discuss several possible sources of systematic errors,
including: sample incompleteness, unknown survey selection function,
uncertainties in the lensing galaxy properties and lens modeling,
in order to verify their effect on the cosmological constraints.

First of all, one general concern is given by the fact that strong gravitational lenses are a biased sample of galaxies.
Most of the previous works found no evidence for a biased sample of the
lensing population with respect to massive early-type galaxies (see, e.g., \citet{Treu06}).
On the other hand, \citet{Faure11} found possible evidence for the
stellar  mass  of lensing early-type galaxies to evolve significantly with redshift.
However, it is still not clear whether this supports a stronger lensing bias
toward massive objects at high redshift or if it is a consequence of the possible higher proportion
of massive and high stellar density galaxies at high redshift.
This could be addressed in dedicated numerical simulations \citep{van09,Man09},
as the available lens samples cannot allow yet to discriminate between the
two alternatives.

We now estimate the systematic errors due to statistical sample incompleteness.
As both our Samples (A, B) have been put together from different surveys,
differences in the observing strategies (and selection functions) may cause
systematical errors hard to estimate. In order to evaluate the effects due to a selection bias,
we have rederived the best estimate on $\Omega_\Lambda$
(with a no-evolving lens population)
by using the whole catalog of $n=122$ lenses and two additional, smaller samples (see Table~\ref{subsample}):
the first one includes only galaxies from the SLACS,
and the second one the gravitational lenses with separations
not larger than $2''$, including 59 and 51 systems, respectively.
All of these three samples clearly suffer from strong selection effects or
are very inhomogeneous
and are therefore not suitable for deriving
constraints by means of the redshift distribution test.
However, they can shed light on the amplitude of the possible systematic errors
due to sample inhomogeneity, incompleteness, and selection bias toward non-representative systems.
For instance, it is well known that the SLACS catalog is characterized by
a selection function favoring moderately massive ellipticals (e.g., \citet{Arneson12});
also the somehow extreme case of Sample C is not complete (below $\Delta \theta = 2''$) as
the probability to detect a lensing galaxy is related to the image separation
as very small-separation lenses easily escape detection
(both in present-day imaging and spectroscopic surveys).
Therefore, we expect that the estimates of $ \Omega_\Lambda$
obtained from these samples allow us to establish an upper limit
on the systematic errors due to a not well-defined lens sample.

\begin{figure}
   \centering
  \includegraphics[width=0.5\textwidth]{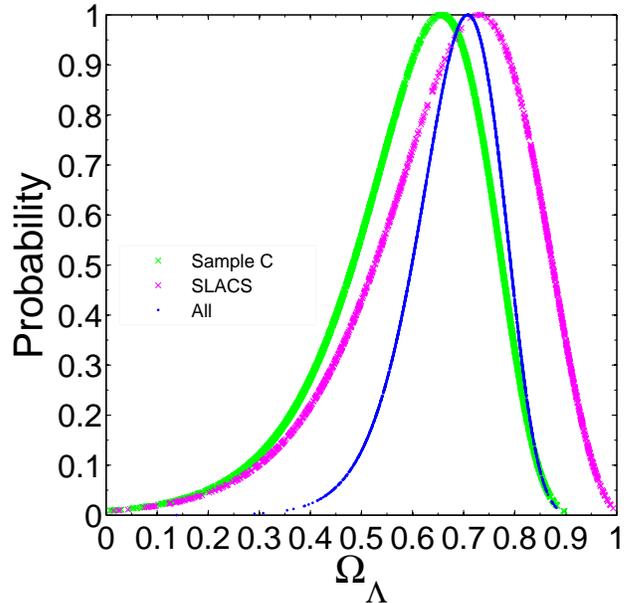}
\caption{Normalized likelihood plot
    as a function of $\Omega_\Lambda$ for the flat $\Lambda$CDM model, for the three biased samples
    described in the text.}
\label{L1}
\end{figure}

Results are shown in Figure~\ref{L1}.
When using the apparently incomplete sample of small separation lenses (Sample C) we find
$ \Omega_\Lambda= 0.65^{+0.11}_{-0.15}$.
This rather smaller value can be related to the fact  that lower-mass lenses,
producing small-separation images,
will tend to be located at redshifts lower than expected
in a large-$\Omega_{\Lambda}$ model.
Hence, this determines a slightly lower value for the cosmological constant.

When adopting the whole SLACS sample, we find $\Omega_\Lambda= 0.73^{+0.14}_{-0.18} $.
Finally, by considering the whole, inhomogeneous list of 122 lenses,
we obtain $ \Omega_\Lambda = 0.71^{+0.07}_{-0.08}$.
When we compare these values with the best estimate obtained from Sample A,
we note that systematic errors do not exceed $\sim 0.1$ on the cosmological constant.

We have assumed all the lenses to be isolated systems,
with negligible line-of-sight contamination.
However, current studies find that proximate
galaxies \citep{Cohn04} and environmental groups \citep{Keeton04}
can have various effects on the primary lens galaxies, even though the
most significant effects only appear to be biasing the galaxy ellipticities and image multiplicities.
Nevertheless, most of lenses in Sample B come from the SLACS survey where the
role of environment has been assessed in \citet{Treu09}.
Namely, it was found that for SLACS lenses the
typical contribution from external mass distribution is no more than a few percent.
Therefore, the environmental effects on
observed image separations appear to be relatively small
(certainly smaller than statistical errors arising from the current sample size of lenses).

In addition to the main lens galaxy, the contribution from line-of-sight density
fluctuations to the lens potential should also be taken into consideration.
Based on the final lens sample from SQLS, \citet{Oguri12} have investigated the line-of-sight effect
and found that its effect on the total lensing probability is rather small compared with the
contribution of other systematic errors to the systematic
error on $\Omega_{\Lambda}$ (see Table~2 in \citet{Oguri12}).
Meanwhile, it is noted that this effect may also have a larger impact
on lensing probabilities for the lens samples with larger images separations \citep{Oguri05,Faure09}.

\begin{figure}
   \centering
   \includegraphics[width=0.5\textwidth]{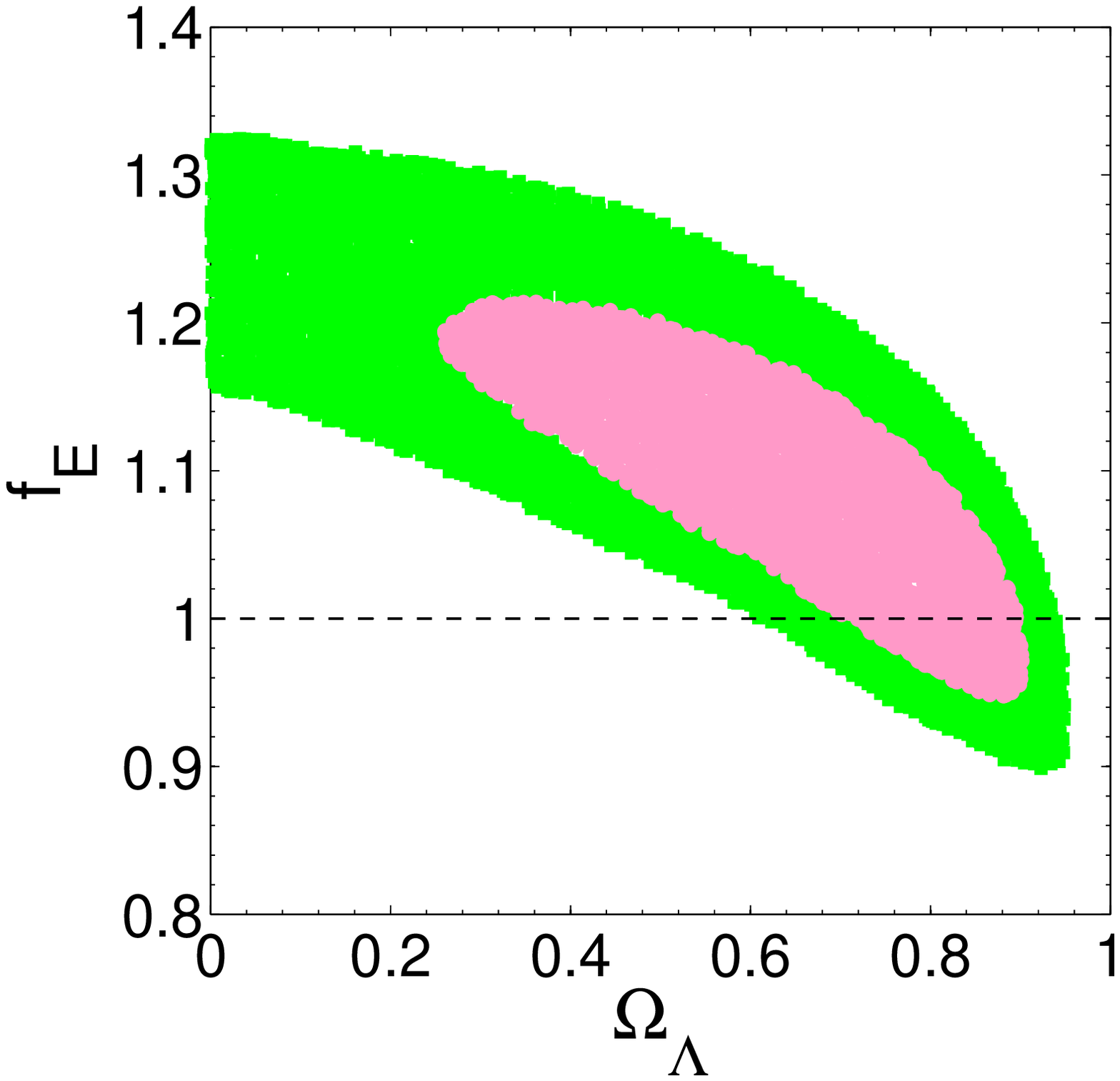}
   \includegraphics[width=0.5\textwidth]{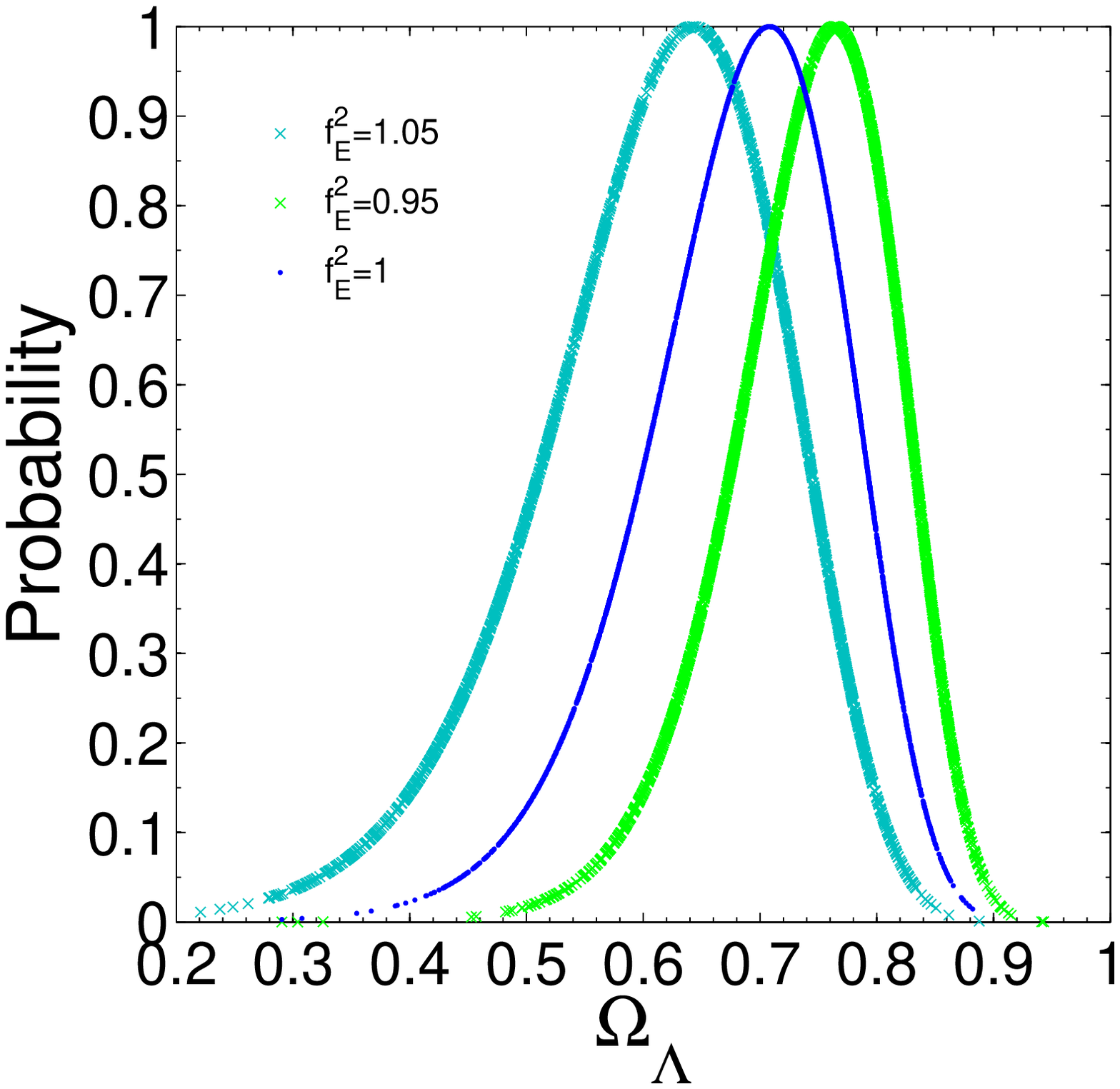}
\caption{Upper panel: constraint results on the cosmological constant $\Omega_\Lambda$ and the parameter $f_E$ for the flat $\Lambda$CDM model with the full sample ($n=122$ lenses). The horizontal dashed line indicates the fiducial value $f_E=1$ assumed in the paper.
Lower panel: constraints on $\Omega_\Lambda$ with different values of $f_E$ considering a fiducial error of $\sim 5\%$ on the values of image separations.}
\label{feerror}
\end{figure}

\begin{figure}
   \centering
   \includegraphics[width=8.9cm]{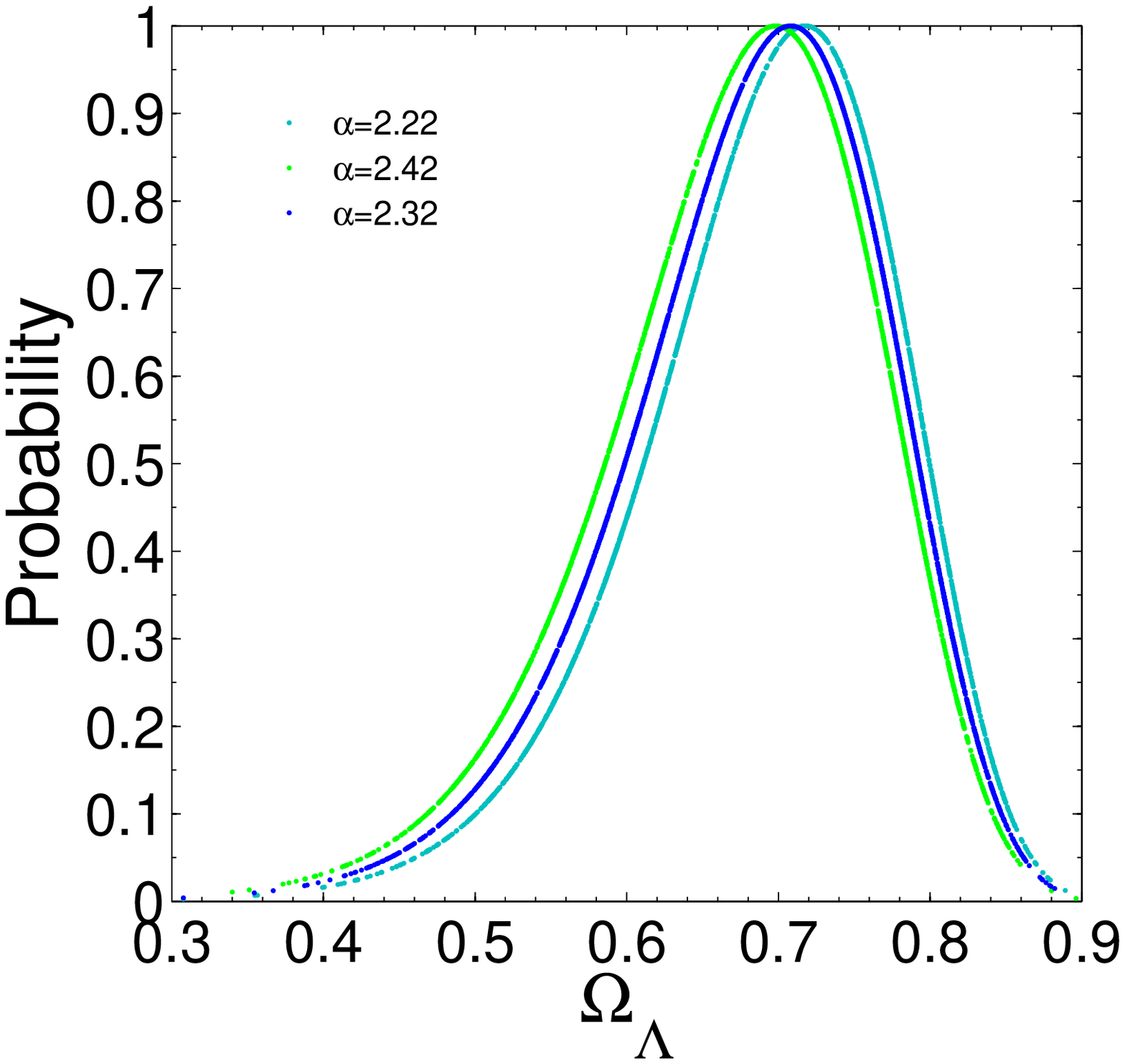}
    \includegraphics[width=8.9cm]{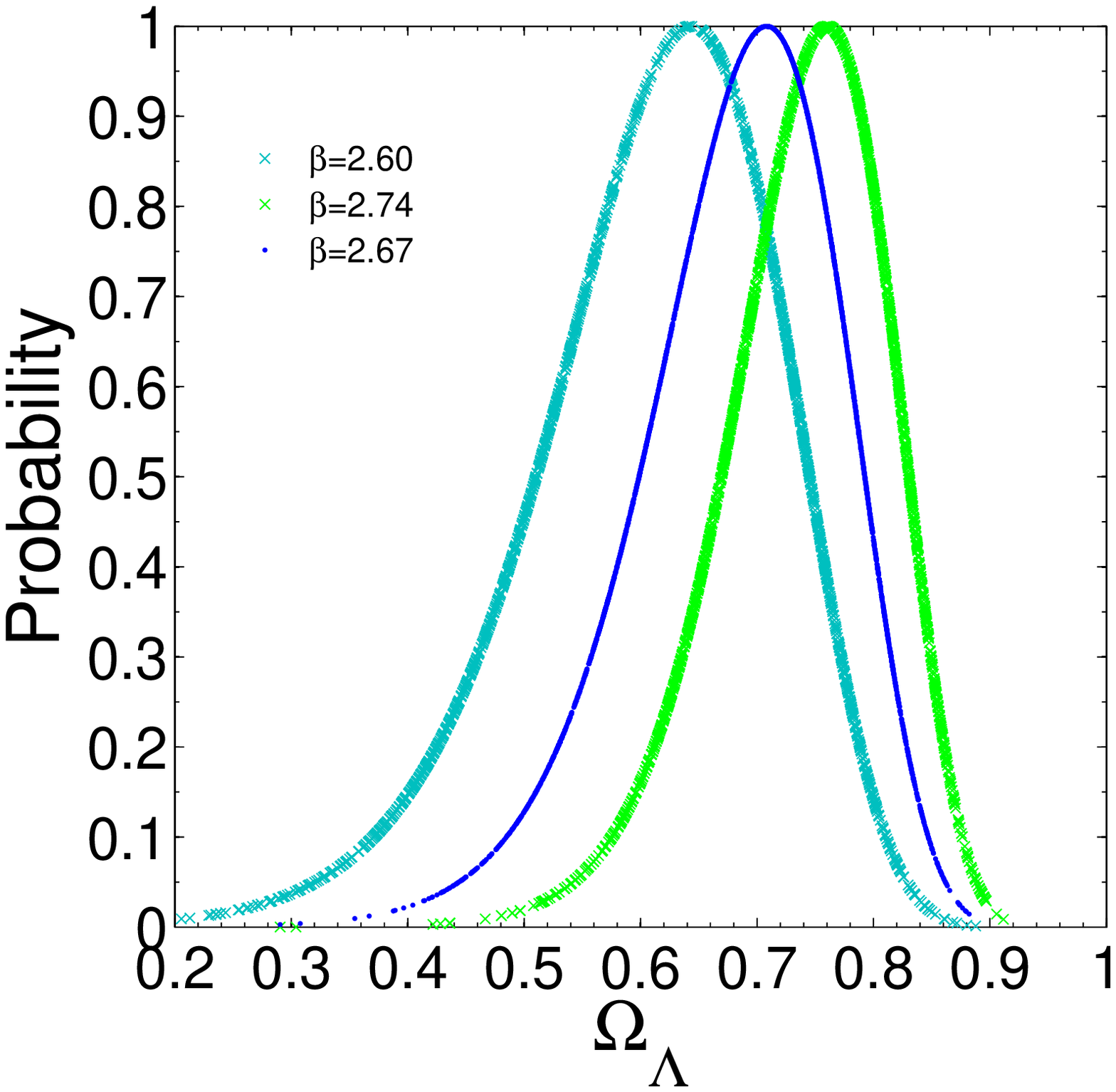}
\caption{Normalized likelihood plot for the flat $\Lambda$CDM model with the full sample ($n=122$ lenses), by introducing the uncertainties on the VDF parameters: the faint-end slope $\alpha$ and the high-velocity cut-off $\beta$, as listed in Table~\ref{prior}.}
\label{VDFerror}
\end{figure}

Evolution of the VDF and model uncertainties can introduce additional systematic errors.
Here we estimate these systematic errors on the constraint results of the flat $\Lambda$CDM with the full sample ($n=122$ lenses),
in a similar way as done in \citet{Oguri12}.
The analysis in Section~\ref{sec:result} suggests that unconstrained redshift evolution of the
velocity function is one of the most significant sources of systematic error.
An additional source of uncertainty is the relation between velocity dispersions and image separations.
This uncertainty is not only related to the difference between
the velocity dispersion $\sigma_{SIS}$ of the mass distribution and
the observed stellar velocity dispersion $\sigma_{0}$ \citep{White96},
but also many complexities such as the velocity dispersion normalization
factor for non-spherical galaxies \citep{Oguri12} and the detailed luminosity profiles of galaxies.

Hence, we introduce the parameter $f_{E}$ that combines the velocity dispersion
$\sigma_{SIS}$ and the stellar velocity dispersion $\sigma_0$
\begin{equation}
\sigma_{SIS} \, = \, f_{E} \sigma_{0},
\label{f_E}
\end{equation}
The Einstein radius given by Equation~\ref{theta1} is then modified as
\citep{Kochanek92,Ofek03}:
\begin{equation}
 \Delta \theta_{*} = 8\pi \left( \frac{\sigma_{*}}{c} \right)^{2} \frac{D_{\rm ls}}{D_{\rm s}} f_{E}^2 \, .
\label{theta2}
\end{equation}
More specifically, we have kept $f_{E}$ as a free parameter to mimic the following effects of:
(1) systematic errors in the difference between $\sigma_{0}$ and $\sigma_{SIS}$;
(2) the influence of line-of-sight mass contamination \citep{Keeton97}; (3) softened isothermal sphere model which may change the typical image separations \citep{Narayan96}; and (4) the effect of secondary lenses on observed image separation. Recently, by combining stellar kinematics (central velocity dispersion measurements) with lensing geometry (Einstein radius determination from position of images), \citet{Cao12b} have tested a combined gravitational lens data set including 70 data points from SLACS and LSD, and obtained constraints on $f_E$ consistent with the previous results \citep{Ofek03}.

We consider the flat $\Lambda$CDM model and obtain simultaneous constraints on $\Omega_\Lambda$ and $f_E$ with the full sample ($n=122$ lenses).
The constraint result is shown in Figure~\ref{feerror}, with the marginalized constraints $f_E=1.08 \pm 0.14$
and $\Omega_\Lambda=0.65^{+0.35}_{-0.45}$. It is obvious that the cosmological constraint on $\Omega_\Lambda$ becomes much weaker, though the concordance $\Lambda$CDM is still favored within $1\sigma$ error region. A vanishing cosmological constant is still ruled at 1$\sigma$.

The slightly larger value of $f_E$ indicates that larger $\Omega_\Lambda$ is preferred by the lens statistics with $f_E=1$ used in this paper. This tendency is consistent with the previous result of \citet{Oguri12} and may help explain why the observed image separation seems to be higher than model predictions \citep{CN07}.  However, we find that the $f_E=1$ case still consists with the data better to than 2$\sigma$. To be more specific, in order to take into account the measurement uncertainty and the approximations we are doing (no lens ellipticity accounted and no external shear), we estimate a fiducial error of $\sim 5\%$ on the values of image separations \citep{Grillo08}, which is equivalent to a $\sim 5\%$ uncertainty on the parameter $f_E^2$.

The velocity distribution function given by Equation~\ref{stat2} is another important source of systematic error on the final results. While adopting the best-fit values of the VDF measurement
in the SDSS Data Release 5 by \citet{Choi07} as our fiducial model, we investigate how the cosmological results are altered by introducing the uncertainties on $\alpha$ and $\beta$ as listed in Table~\ref{prior}. With the results of measurements on the VDF, we vary the parameter of interest while fixing the other parameters at their best-fit values.
For example, based on the $n=122$ sample, we vary the faint-end slope $\alpha$ by $\pm 0.10$ and find that this effect is quite negligible when compared to the present accuracy of the test \citep{Mit05}.

The complete set of standard priors and allowances included in the analysis of the above systematics is summarized in Table~\ref{prior}. As shown in Figure~\ref{feerror} and \ref{VDFerror} and Table~\ref{priorresult}, by comparing their contribution to the systematic error on $\Omega_\Lambda$ for the flat $\Lambda$CDM with the full sample, we find that the largest sources of systemic error are the dynamical normalization $f_E$ and the high-velocity cutoff $\beta$, followed by the faint-end slope $\alpha$ of the VDF. This finding is consistent with the earlier results in \citet{Oguri12}.
Meanwhile, this result is also compatible to the concordance cosmological
model ($\Omega_{\Lambda} \sim 0.7$, and $\Omega_{m}\sim 0.3$).
Indeed, current samples of lenses do not allow to discriminate between an $\Omega_\Lambda$ and a dynamical dark energy component.

\begin{table}
\caption{\label{prior} Summary of the Priors and Standard Systematic Allowances Included in the Analysis.}
\begin{center}
\begin{tabular}{c|c}\hline\hline
 Parameter & \hspace{4mm}Allowance\hspace{4mm}\\ \hline

 EoS of DE  ($w$) & $-5<w<0$ \\
 DE density in XCDM ($\Omega_x$) & $0<\Omega_x<1$ \\
 DE density in $\Lambda$CDM ($\Omega_\Lambda$) & $0<\Omega_\Lambda<1$ \\
 Matter density ($\Omega_m$) & $0<\Omega_m<1$ \\
 Evolution of $n_{*}$ ($P$) & $-10<P<10$ \\
 Evolution of $\sigma_{*}$ ($U$) & $-1<U<1$ \\
 Normalization factor ($f_E$) &$(0.5)^{1/2}<f_{E}<(2.0)^{1/2}$\\
 Faint-end slope ($\alpha$) & $\alpha=2.32 \pm 0.10$\\
 High-velocity cut-off ($\beta$) & $\beta = 2.67 \pm 0.07$ \\
 \hline\hline
\end{tabular}
\end{center}

\end{table}

\begin{table}
\caption{\label{priorresult} Constraint Results Obtained by the Full Sample ($n=122$ lenses) for the Flat $\Lambda$CDM Model with Different Systematic Errors.}
\begin{center}
\begin{tabular}{c|c}\hline\hline
 Systematics & \hspace{4mm}$\Omega_\Lambda$\hspace{4mm}\\ \hline
$f_E^2=1.00;\alpha=2.32;\beta=2.67$  & $\Omega_\Lambda= 0.71_{-0.08}^{+0.07}$     \\
$f_E^2=0.95;\alpha=2.32;\beta=2.67$ & $\Omega_\Lambda= 0.76_{-0.07}^{+0.06}$     \\
$f_E^2=1.05;\alpha=2.32;\beta=2.67$ & $\Omega_\Lambda= 0.65_{-0.12}^{+0.08}$     \\
$f_E^2=1.00;\alpha=2.22;\beta=2.67$ & $\Omega_\Lambda= 0.72_{-0.09}^{+0.08}$     \\
$f_E^2=1.00;\alpha=2.42;\beta=2.67$ & $\Omega_\Lambda= 0.70_{-0.10}^{+0.08}$     \\
$f_E^2=1.00;\alpha=2.32;\beta=2.60$ & $\Omega_\Lambda= 0.64_{-0.10}^{+0.08}$     \\
$f_E^2=1.00;\alpha=2.32;\beta=2.74$ & $\Omega_\Lambda= 0.76\pm 0.07$     \\
 \hline\hline
\end{tabular}
\end{center}

\end{table}

\section{Conclusion and discussion }
\label{sec:conclusion}

Since the discovery of the accelerating expansion of the universe,
in addition to the standard $\Lambda$CDM cosmological model, a large
number of theoretical scenarios have been proposed for the acceleration mechanism.
Examples include the quintessence
\citep{Ratra88,Caldwell98}, phantom
\citep{Caldwell02}, quintom \citep{Feng05,Feng06,Guo05}
and the Chaplygin gas \citep{Kamenshchik01,Bento02,Alam03,Zhu04,Zhang06}.
All these acceleration mechanisms should be tested with various astronomical
observations such as SNeIa, WMAP \citep{Komatsu09}
and BAO  \citep{Percival07}.
However, it is still important to use
as many as different observational probes to set constraints on
the cosmological parameters.
In this work, we have followed this
direction and used the redshift distribution of
two well-defined samples of lensing, elliptical galaxies
drawn from a large catalog of 122 gravitational lenses
from a  variety of surveys (see Table~\ref{tab:data}). The BIC is also applied in this analysis to assess various dark energy models and make
a comparison.

Considering the two cases with and without the redshift evolution of the velocity function of galaxies,
we have analyzed four dark energy models
(the $\Lambda$CDM, the XCDM with constant $w$, the DGP and the RDE models)
under a flat universe assumption.
For each model, we have calculated the best-fit values of its parameters and found its $\Delta{\rm BIC}$ value.
Results are plotted in Figure~\ref{L}-\ref{RDE}.

The fit and information criteria results are summarized in Table~\ref{tab:result}.
It is shown that for the zero-curvature $\Lambda$CDM model, the likelihood is maximized,
for $\Omega_\Lambda= 0.70 \pm 0.09$
with no redshift evolution and $\Omega_\Lambda= 0.73\pm 0.09$ with redshift evolution,
when using the Sample A (see Section~2). Consistent results (within $1\sigma$) are derived when the alternative Sample B is considered.

We have also derived simultaneous constraints on the redshift evolution
of the parameters $n_*$ and $\sigma_*$ of the velocity function.
The constraints in the $P-U$ plane also indicate that the data are consistent
with the no-redshift-evolution case ($P=0; \, U=0$)
at $1\sigma$, with $P=-1.2 \pm 1.4$ and $U=0.22_{-0.27}^{+0.26}$.
Both in the no-evolution and galaxy evolution scenario,
we rule out with high confidence a vanishing cosmological constant, with both the lens samples
(at confidence larger than $\sim 4 \sigma$), as recently found by \citet{Oguri12}
by adopting a smaller and independent lens sample.
Therefore, the redshift test adds an independent evidence
to the accelerating expansion.

The obtained likelihood distributions shown in Figure~\ref{L} are also in agreement with
the results from analyzing data of WMAP 5-year results with the BAO and SNe Union data,
and the large-scale structures in the SDSS
luminous red galaxies \citep{Spe03,Tegmark04,Eisenstein05},
which implies that gravitational lensing statistics provides an
independent and complementary support on the $\Lambda$CDM model.

\begin{figure}
   \centering
   \includegraphics[width=8.9cm]{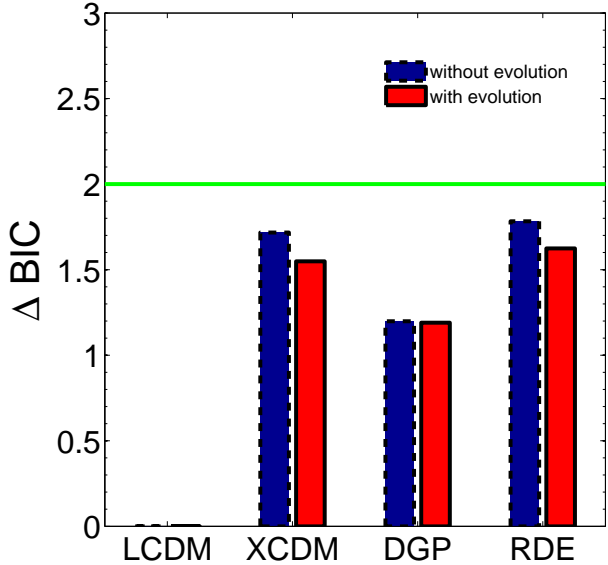}
\caption{ Graphical representation of the values of $\Delta$BIC for each
model, relative to the LCDM ($\Lambda$CDM). The horizontal green line indicates $\Delta$BIC=2, which is considered positive evidence against the model with the higher BIC.
\label{deltaB}}
\end{figure}

We give a graphical representation of the BIC test Figure~\ref{deltaB}.
Following the $\Lambda$CDM model,
the DGP model is the only one-parameter model that reduces to the $\Lambda$CDM but gives a worse fit,
as we obtained close values for the matter density.
The XCDM model gives a comparably good fit,
but has one additional free parameter, that is in accordance with the best-fit $\Lambda$CDM model
(within 1$\sigma$ range for the EoS parameter).
The other two-parameter model, RDE, provides a worst fit to the data,
though the difference in BIC ($\Delta{\rm BIC}$) indicates no
clear evidence against it.
Therefore, while still not firmly ruling out
competing world models, the redshift distribution test
clearly favors the cosmological constant model, a conclusion in accordance with previous works \citep{Davis07}.

\begin{table}
\caption{\label{tab:result} Summary of the constraint results and information criterion with Sample A.
The $\Delta$BIC values for all other models are
measured with respect to the $\Lambda$CDM model.}
\begin{tabular}{c|l|llllllllllllllll}\hline\hline
Model        & Constraint result &   $\Delta$BIC \\
\hline
$\Lambda$CDM ($P=U=0$)  & $\Omega_\Lambda= 0.70\pm 0.09$      & 0  \\
 ~~~~~~~~~~~($P\neq U\neq0$)  & $\Omega_\Lambda= 0.73\pm 0.09$      & 0  \\
  XCDM ($P=U=0$)  & $\Omega_x= 0.77\pm 0.17$; $w= -2.3_{-2.7}^{+1.3}$      & 1.72  \\
 ~~~~~~~~~~~($P\neq0 \neq0$)  & $\Omega_x= 0.79\pm 0.13$; $w= -2.1_{-2.8}^{+1.1}$      & 1.55 \\
  DGP ~~ ($P=U=0$)  & $\Omega_m= 0.25_{-0.09}^{+0.11}$      & 1.20   \\
~~~~~~~~~~~($P\neq U\neq0$)  & $\Omega_m= 0.22_{-0.09}^{+0.10}$      & 1.19  \\
  RDE ~~ ($P=U=0$)  & $\Omega_m=0.22^{+0.10}_{-0.11}; \beta =0.29\pm 0.19$      & 1.78  \\
~~~~~~~~~~($P\neq U\neq0$)  & $\Omega_m=0.18^{+0.11}_{-0.12}; \beta =0.28\pm 0.18$      &  1.62   \\

\hline
\end{tabular}
\end{table}

In order to asses the reliability of our results and
the related systematic errors,
we have introduced three additional lens samples,
characterized by different degrees of inhomogeneity and incompleteness,
and re-derived an estimate for the cosmological constant assuming a not evolving lens population.
Two lens samples (the whole heterogeneous catalog of 122 systems and the SLACS sample)
give results fully consistent with those discussed above,
confirming us that systematic errors due to sample selection are not larger than
statistical uncertainties.

Our model involves several uncertainties and assumptions
that introduce additional systematic errors in our cosmological analysis (see Table~\ref{prior}).
By comparing the contribution of each of these systematic errors to the systematic error on the flat $\Lambda$CDM model (see Figure~\ref{feerror} and \ref{VDFerror} and Table~\ref{priorresult}), we find that the largest sources of systematic
error are the dynamical normalization and the high-velocity cutoff factor, followed by the faint-end slope of the velocity dispersion function, which is consistent with the earlier results in \citet{Oguri12}. Moreover, the comparable systematic errors suggest the importance of careful studies of the systematics for robust cosmological constraints from lens statistics.

Finally, we note that four important effects should be mentioned.
Evolution of the source population can also matter the technique applied in this paper, which might have a small second-order effect on the statistics \citep{Oguri12}.
We have also neglected systematic uncertainties due to the effect of small-scale inhomogeneities on large-scale observations.
In fact, the inhomogeneous matter distribution can affects light propagation
and the related cosmological distances \citep{Sereno01,Sereno02b}, although its effect on the total
lensing statistics is small \citep{Covone05}.
Meanwhile, the lens redshift test applied
in this paper may also lose the statistical power of absolute lensing probabilities.
The last one is, multiple errors or biases in the method could easily be canceled out, which may lead the result to be a statistical fluke \citep{Maoz05}.

Despite some of its inherent difficulties, the redshift distribution test,
with either larger gravitational lensing samples from future wide-field surveys such as Pan-STARRS and Large Synoptic Survey Telescope by taking advantage of time-domain information \citep{Oguri10} or a joint
investigation with other cosmological observations, could be helpful for advancing
such applications and provide more stringent constraints on the cosmological parameters.

\vspace{0.5cm}

We acknowledge fruitful discussions with M. Paolillo.
We thank the anonymous referee for his/her valuable comments which helped us to improve the paper.
GC thanks I. Skripnikova for enlightening discussions.
This work was supported by the National Natural Science Foundation
of China under the Distinguished Young Scholar Grant 10825313 and
Grant 11073005, the Ministry of Science and Technology national
basic science Program (Project 973) under Grant No. 2012CB821804, the
Fundamental Research Funds for the Central Universities and
Scientific Research Foundation of Beijing Normal University, and the Excellent Doctoral
Dissertation of Beijing Normal University Engagement Fund.

\end{document}